\documentclass[conference,letterpaper]{IEEEtran}
\usepackage[utf8]{inputenc} 
\usepackage[T1]{fontenc}
\usepackage{url}
\usepackage{ifthen}
\usepackage{cite}
\usepackage[cmex10]{amsmath} %

\usepackage{amsfonts} %
\usepackage[caption=false,font=normalsize,labelfont=sf,textfont=sf]{subfig}
\usepackage{graphicx}

\usepackage{tikz}
\usepackage{pgfplots}
\usepackage{pgfplotstable}
\pgfplotsset{compat=newest}

\usepackage[detect-all]{siunitx}
\usepackage{booktabs}
\usepackage{circuitikz}
\usepackage{mathtools}
\usepackage{hyperref}
\hypersetup{unicode=true,pdfborder = {2 0 0.1}}

\usetikzlibrary{arrows,shapes.misc,chains,scopes,shadows,arrows.meta}
\usetikzlibrary{calc}
\usetikzlibrary{fit}
\usetikzlibrary{intersections}
\usetikzlibrary{shapes}
\usetikzlibrary{scopes}
\usetikzlibrary{backgrounds}
\usetikzlibrary{positioning,decorations.pathreplacing,calc}
\usetikzlibrary{decorations.text}

\DeclareSIUnit\bpcu{bpcu}
\DeclareSIUnit\bpQs{bpQs}

\newcommand{\bm}[1]{\mathbf{#1}}

\newcommand{\sicindex}{t} 
\newcommand{\siclength}{N}

\newcommand{\lay}{\ell}
\newcommand{\maxlay}{L}

\newcommand{\osidx}{u}

\pgfmathsetmacro{\blockwidth}{30}
\pgfmathsetmacro{\blockheight}{16}

\tikzstyle{comblock} = [drop shadow={shadow xshift=0.1em,shadow yshift=-0.1em},thick,black,fill=white,draw=black, rectangle,minimum height=\blockheight pt,minimum width=\blockwidth pt]

\tikzstyle{sumstyle} = [thick,draw, circle,inner sep=-1pt,outer sep=0pt,font=\normalsize,fill=white,drop shadow={shadow xshift=0.1em,shadow yshift=-0.1em},]

\tikzstyle{comtriag} = [regular polygon, regular polygon sides=3,
              draw=black, black,  text width=0.90em,
              inner sep=0.9mm, outer sep=0mm,
              shape border rotate=-90]

\tikzstyle{input} =  [coordinate]
\tikzstyle{output} = [coordinate]

\DeclareMathOperator*{\T}{T}

\DeclareMathOperator{\nonl}{\xi}
\DeclarePairedDelimiter\parens{\lparen}{\rparen}
\newcommand{\nonlp}[1]{\nonl\parens*{#1}}

\DeclareMathOperator*{\argmin}{argmin}

\DeclareMathOperator*{\sinc}{sinc}
\DeclareMathOperator*{\vstack}{cat}

\renewcommand{\tt}{\text{t}}

\newcommand{\dimC}[1]{\ensuremath{\in \mathbb{C}^{#1}}}
\newcommand{\dimR}[1]{\ensuremath{\in \mathbb{R}^{#1}}}

\usetikzlibrary{intersections}
\usetikzlibrary{calc}
\usetikzlibrary{shapes}
\usetikzlibrary{scopes}
\usetikzlibrary{backgrounds}
\usetikzlibrary{positioning,decorations.pathreplacing}

\definecolor{TUMBlack}{cmyk}{0,0,0,1}     %
\definecolor{TUMWhite}{cmyk}{0,0,0,0}     %

\definecolor{TUMBlue} {cmyk}{1,0.43,0,0}  %

\definecolor{TUMDarkBlue}   {cmyk}{1,0.57,0.12,0.7}      %
\definecolor{TUMDarkerBlue} {cmyk}{1,0.54,0.04,0.19}     %
\definecolor{TUMMediumBlue} {cmyk}{0.9,0.48,0,0}         %
\definecolor{TUMLighterBlue}{cmyk}{0.65,0.19,0.01,0.04}  %
\definecolor{TUMLightBlue}  {cmyk}{0.42,0.09,0,0}        %

\definecolor{TUMDarkGray}  {cmyk}{0,0,0,0.8}  %
\definecolor{TUMMediumGray}{cmyk}{0,0,0,0.5}  %
\definecolor{TUMLightGray} {cmyk}{0,0,0,0.2}  %

\definecolor{TUMGreen} {cmyk}{0.35,0,1,0.2}         %
\definecolor{TUMOrange}{cmyk}{0,0.65,0.95,0}        %
\definecolor{TUMIvory} {cmyk}{0.03,0.04,0.14,0.08}  %

\definecolor{TUMBeamerYellow}    {rgb}{1.00,0.71,0.00}  %
\definecolor{TUMBeamerOrange}    {rgb}{1.00,0.50,0.00}  %
\definecolor{TUMBeamerRed}       {rgb}{0.90,0.20,0.09}  %
\definecolor{TUMBeamerDarkRed}   {rgb}{0.79,0.13,0.25}  %
\definecolor{TUMBeamerBlue}      {rgb}{0.00,0.60,1.00}  %
\definecolor{TUMBeamerLightBlue} {rgb}{0.25,0.75,1.00}  %
\definecolor{TUMBeamerGreen}     {rgb}{0.57,0.67,0.42}  %
\definecolor{TUMBeamerLightGreen}{rgb}{0.71,0.79,0.51}  %

\definecolor{I8LogoRed}         {rgb}{0.51,0,0.08}
\definecolor{I8LightBlue}       {rgb}{0.725,0.812,0.882}
\definecolor{I8DarkBlue}        {rgb}{0.490,0.573,0.667}
\definecolor{I8Blue}            {rgb}{0.576,0.624,0.718}

\definecolor{mycolor1bright}{HTML}{E9D8A6}%
\definecolor{mycolor7bright}{HTML}{EE9B00}%
\definecolor{mycolor6bright}{HTML}{AFCBFF}%
\definecolor{mycolor5dark}{HTML}{166337}%
\definecolor{mycolor6}{HTML}{FF0000}%

\definecolor{mycolor6dark}{HTML}{961717}%
\definecolor{mycolorlightorange}{HTML}{7A7A7A}%
\definecolor{mycolorcube}{HTML}{C2C2C2}%
\definecolor{mycolor5bright}{HTML}{FFFFFF}%

\DeclareFontEncoding{LS1}{}{}
\DeclareFontSubstitution{LS1}{stix}{m}{n}
\DeclareSymbolFont{stixletters}{LS1}{stix}{m}{it}
\DeclareMathAccent{\cev}{\mathord}{stixletters}{"91}
\DeclareMathAccent{\vec}{\mathord}{stixletters}{"92}

\newcommand{\fw}[1]{\vec{#1}} 
\newcommand{\bw}[1]{\cev{#1}}

\newcommand{\Wfw}[2]{\fw{W}_{#1}^{#2}} %
\newcommand{\Cfw}[2]{\fw{C}_{#1}^{#2}} %
\newcommand{\Cbw}[2]{\bw{C}_{#1}^{#2}} %
\newcommand{\hfw}[2]{\fw{\mathbf{h}}_{#1}^{#2}} %
\newcommand{\hbw}[2]{\bw{\mathbf{h}}_{#1}^{#2}} %

\newlength\nntablevspace
\makeatletter
\DeclareRobustCommand{\rvdots}{%
  \vbox{
    \baselineskip4\p@\lineskiplimit\z@
    \kern-\p@
    \hbox{.}\hbox{.}\hbox{.}
  }}
\makeatother

\begin{document}
\title{Neural Network Equalizers and\\
Successive Interference Cancellation for\\ Bandlimited Channels with a Nonlinearity}

\author{%
\IEEEauthorblockN{Daniel Plabst\IEEEauthorrefmark{1}, 
Tobias Prinz\IEEEauthorrefmark{1}, Francesca Diedolo\IEEEauthorrefmark{1}, Thomas Wiegart\IEEEauthorrefmark{1},\\ Georg Böcherer\IEEEauthorrefmark{2}, Norbert Hanik\IEEEauthorrefmark{1} and Gerhard Kramer\IEEEauthorrefmark{1}}
\IEEEauthorblockA{\IEEEauthorrefmark{1}School of Computation, Information and Technology, Technical University of Munich, Germany%
}
\IEEEauthorblockA{\IEEEauthorrefmark{2}Huawei Munich Research Center, Germany
}
}

\maketitle

\begin{abstract}
Neural networks (NNs) inspired by the forward-backward algorithm (FBA) are used as equalizers for bandlimited channels with a memoryless nonlinearity. The NN-equalizers are combined with successive interference cancellation (SIC) to approach the information rates of joint detection and decoding (JDD) with considerably less complexity than JDD and other existing equalizers. Simulations for short-haul optical fiber links with square-law detection illustrate the gains. %
\end{abstract}

\section{Introduction}
\label{sec:introduction}
\IEEEPARstart{C}{ommunication} systems with hardware constraints or high transmit power may introduce non-linearities, e.g., via a power amplifier (PA)~\cite{mollen2016massive} or a square-law detector (SLD)~\cite{chagnon_optical_comms_short_reach_2019}. Non-linearities degrade performance in general, and one may need joint detection and decoding (JDD) to approach capacity. JDD is often too complex~\cite{muller_capacity_separate2004}, and the complexity can be reduced by separate detection and decoding (SDD) for each channel input symbol; see~\cite{sheik_achievable_2017,liga_information_2017}. However, SDD rates may be significantly lower than JDD rates~\cite{muller_capacity_separate2004,prinz2023successive}. Two methods that use SDD to approach JDD performance are turbo detection and decoding (TDD)~\cite{douillard1995iterative,alexander-ETT98,wang1999iterative} and multi-level coding (MLC) with successive interference cancellation (SIC)~\cite{prinz2023successive,wachsmann_multilevel_1999,PfisterAIRFiniteStateChan2001,soriaga_determining_2007}.
TDD requires dedicated code design to approach capacity while MLC-SIC permits using off-the-shelf codes; see~\cite{PfisterAIRFiniteStateChan2001,soriaga_determining_2007,ten2004design}.

We consider MLC-SIC, for which one computes a-posteriori probabilities (APPs), e.g., symbol-wise APPs via the forward-backward algorithm (FBA)~\cite{bcjr_1974}. Two related methods are (soft-output) Viterbi equalization~\cite{hagenauer1989SOVA} and Gibbs sampling (GS)~\cite{mackay2003information} to approximate APPs. 
Linear~\cite{muller_capacity_separate2004} or non-linear equalizers, such as decision feedback equalizers~\cite{cioffi1995mmsedfe} and Volterra filters~\cite{taiho1985secondvolterra} may estimate APPs. The main limitation is that, for high rates, either the complexity is high~\cite{plabst2022achievable,taiho1985secondvolterra} or residual interference reduces rates significantly~\cite{prinz2023successive,muller_capacity_separate2004,plabst2022achievable}.

We use MLC-SIC with neural network (NN) APP detectors and show one can efficiently achieve high rates for large-memory models. We refer to~\cite{plabst2024neural} for a detailed overview of NNs for SDD, JDD and TDD. The literature on MLC-SIC with NNs performs soft-SIC for multiple access~\cite{minghuang2000successive,lin2019deep,kang2020deep,aref2020deep,kim2023enhancedsoft,vanluong2022deep} but with no channel decoding between the stages. Thus, significant error propagation may occur if the interference is not entirely suppressed. Also, the information rate is that of SDD.

This paper shows how to approach JDD performance with MLC-SIC and one NN detector per stage. The NN structure is inspired by the FBA, and, in contrast to the NN literature, the receiver decodes between stages and approaches JDD performance as the number of SIC stages increases. To illustrate the gains, we study short-reach fiber links with long memory and an SLD and compare with mismatched SIC equalizers~\cite{prinz2023successive}. The NN-SIC receiver outperforms and is substantially simpler than existing receivers that approximate JDD.

This paper is organized as follows. Sec.~\ref{sec:system-model} introduces the system model with a non-linearity. Sec.~\ref{sec:detection_and_decoding} reviews the SDD and SIC rates. Sec.~\ref{sec:app_nn} describes the NN-APP equalizer. Sec.~\ref{sec:numerical} presents simulation results, and Sec.~\ref{sec:conclusions} concludes the paper.
 
\emph{Notation:}
Column vectors and matrices are written in bold letters, e.g., $\bm{a}$. The transpose of $\bm{a}$ is $\smash{\bm{a}^{\smash{\T}}}$ and $\vstack{(\mathbf{a},\mathbf{b})} = [ \mathbf{a}^{\T}, \mathbf{b}^{\T}]^{\T}$ stacks $\mathbf{a}$ and $\mathbf{b}$. 
The phase of a complex number $z$ is $\angle z$. We denote a string of scalars by $x_\kappa^n=(x_\kappa,\ldots,x_n)$ and string of vectors by $\mathbf{X}_\kappa^n= (\mathbf{X}_\kappa,\ldots,\mathbf{X}_n)$ and omit the subscript if $\kappa \!=\! 1$. 
The sinc function is $\sinc(t) = \sin(\pi t)/(\pi t)$. 
The  convolution of $g(t)$ and $h(t)$ is $g(t)*h(t)$ and the energy of $a(t)$ is $\|a(t)\|^2 = \int_{-\infty}^{+\infty} \big| a(t) \big|^2 \mathrm{d}t $.
Entropy, conditional entropy, and mutual information are defined as in~\cite[Chap.~2]{cover1991elementsofIT}, and we measure the quantities in bits.

\section{System Model}
\label{sec:system-model}
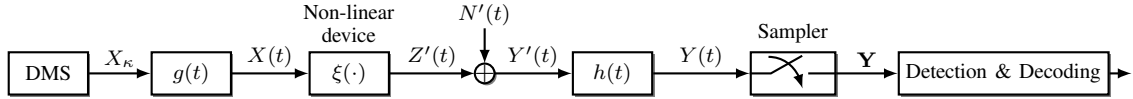
\begin{figure*}[t!]
    \centering
    \usetikzlibrary{decorations.markings}
\tikzset{node distance=2.3cm}

\pgfdeclarelayer{background}
\pgfdeclarelayer{foreground}
\pgfsetlayers{background,main,foreground}
\tikzset{boxlines/.style = {draw=black!20!white,}}
\tikzset{boxlinesred/.style = {densely dashed,draw=red!50!white,thick}}

\pgfmathsetmacro{\samplerwidth}{32}

\tikzset{midnodes/.style = {midway,above,text width=1.5cm,align=center,yshift=-0.1em}}
\tikzset{midnodesRP/.style = {midway,above,text width=1.5cm,align=center,yshift=-1.4em}}

\begin{tikzpicture}[]
    \footnotesize
    \node[comblock] (dms) {DMS};
    \node[comblock,node distance=1.9cm,right of=dms] (txfilter) {$g(t)$};
    \node [comblock,right of=txfilter,node distance=2.1cm] (sld) {$\nonlp{\cdot}$};
    \node [sumstyle,right of=sld,node distance=1.8cm] (sumnode) {$+$};
    \node [input, name=noise,above of=sumnode,node distance=0.6cm] {Input};
    \node [comblock,right of=sumnode,node distance=1.7cm] (rxfilter) {$h(t)$};
    \node [comblock,right of=rxfilter,minimum width=\samplerwidth pt,node distance=2.4cm] (sampler) {};
    \node [comblock,right of=sampler,node distance=2.8cm] (dsp) {Detection \& Decoding};
    \node [input, name=output, right of=dsp,node distance=1.7cm] {Output};
    \draw[thick] (sampler.west) -- ++(\samplerwidth/4 pt,0) --++(\samplerwidth/2.7 pt,\samplerwidth/4.8 pt );
    \draw[thick] (sampler.east) -- ++(-\samplerwidth/3pt,0);

    \draw ($(sampler.west) + (\samplerwidth/4.5 pt,0.22)$)edge[out=0,in=100,-latex,thick] ($(sampler.east) + (-\samplerwidth/2.5 pt,-0.2)$);

    \draw[-latex,thick] (dms) -- node[midnodes](){$X_\kappa$} (txfilter);
    \draw[-latex,thick] (txfilter) -- node[midnodes](s_ti){$X(t)$  } (sld);

    \draw[-latex,thick] (sld) -- node[midnodes](){$Z^\prime(t)$\\[0.2em] } (sumnode);
    \draw[-latex,thick] (sumnode) -- node[midnodes](){$Y'(t)$\\[0.2em] } (rxfilter);
    \draw[-latex,thick] (rxfilter) -- node[midnodes](){$Y(t)$\\[0.2em] } (sampler);
    \draw[-latex,thick] (sampler) --  node[midnodes](){$\mathbf{Y}$\\ [0.2em] } (dsp)  ;
    \draw[-latex,thick] (dsp) --  node[midnodes,xshift=0cm,align=center](){ }(output); %

    \draw[-latex,thick] (noise) -- (sumnode);
    \node[above,yshift=-0.06cm] () at  (noise) {$N'(t)$};

    \node[above,yshift=1.1em] () at (sampler) {Sampler};
    \node[below,yshift=-1.3em, text width=1.9cm, align=center] () at (sampler) {}; %
    
    \node[below,yshift=-1.3em] () at (rxfilter) {};

    \node[above,yshift=+1.1em,text width=2cm,align=center] () at (sld) {Non-linear device};

\end{tikzpicture}
    \vspace{-12pt}
    \caption{Bandlimited channel with a memoryless non-linearity and additive noise.}
    \label{fig:continuous_detailed_system_model}
\end{figure*}

Fig.~\ref{fig:continuous_detailed_system_model} shows a bandlimited channel with a memoryless non-linear device $\nonlp{\cdot}$ and additive noise $N'(t)$ after the non-linearity. 
This model applies to wireless communication with non-linearities at the transmitter due to PAs, mixers, and digital-to-analog converters (DACs), or at the receiver due to low-noise amplifiers (LNAs), mixers, and analog-to-digital converters (ADCs)~\cite{mollen2016massive}. The model also describes fiber-optic communications with non-linearities at the transmitter due to a driver amplifier, DAC, modulator, and optical amplifiers, or at the receiver with a square-law detector (SLD), a single photodiode~\cite{plabst2022achievable}, LNA, and ADC. 
The noise $N'(t)$ might model amplified thermal noise of a radio frequency amplifier~\cite{friis1944noise,rapp1991effects,saleh1981frequency}, or amplified spontaneous emission noise of an erbium-doped fiber amplifier,
or lumped noise of photo-detection~\cite[Sec.~II]{wiener_filter_plabst2020}, the LNA, and the ADC.

\subsection{Continuous Time Model}
\label{sec:time-continuous-model}
A source generates uniformly, independently and identically distributed (u.i.i.d.) symbols $(X_\kappa)_{\kappa \in \mathbb{Z}} = \left( \ldots, X_1, X_2, \ldots \right)$ from the symbol alphabet $\mathcal{A} = \{a_1,\ldots,a_M \}$ where $M = 2^m$. 
After filtering with $g(t)$, the baseband waveform is
\begin{align}
    X(t)  = \sum\nolimits_{\kappa} X_\kappa \cdot g(t-\kappa T_\text{s})
\end{align}
where $B=1/T_s$ is the symbol rate and $g(t)$ collects all linear effects, including the bandwidth limitations of the pulse, non-linear device, and DAC. 
The non-linear device puts out $Z'(t) = \nonlp{X(t)}$ to which noise $N'(t)$ is added. The noise is modeled as a complex white Gaussian random process with two-sided power spectral density (PSD) $N_0/2$ Watts per Hertz per dimension. The signal and noise are filtered by $h(t)$, which may include a receiver bandwidth limitation and anti-aliasing filter matched to the ADC. Colored noise can be modeled by including a whitening filter in $h(t)$.  
The filtered noise $N(t)=N'(t)*h(t)$ is a stationary circularly-symmetric complex Gaussian process with autocorrelation function (ACF) $
\varphi_\text{NN}(\tau) =  N_0 \cdot (h^*(-\tau) *  h(\tau))$.

\subsection{Memoryless Non-Linear Device}
\label{sec:memoryless_nl}
The non-linearity expands the bandwidth, e.g., if $\nonlp{\cdot}$ is a polynomial of degree $d_\xi$ then $Z'(t)$ occupies $d_\xi$ times the bandwidth of $X(t)$~\cite[Sec.~3.1]{zhou2004spectral},~\cite[Thm.~1]{wise1977effect}. 
More generally, the bandwidth of $Z'(t)$ may be unbounded but is bandlimited via $h(t)$. To illustrate, consider two applications.

\subsubsection{Wireless Transmitter with a PA} Consider~\cite{ochiai2013analysis} where
\begin{align}
    Z'(t) = \nonlp{|X(t)|} \cdot \exp(\mathrm{j} \angle X(t))
\end{align}
with a non-linear real-valued function $\nonlp{\cdot}$ that models a solid-state PA that distorts the magnitude; see~\cite[Sec.~3.2]{rapp1991effects}. 

\subsubsection{Optical Fiber Receiver with a SLD}
\label{sec:fiber_optic_application}
Consider a single polarization and the model from~\cite[Fig.~2]{plabst2022achievable} where $g(t) \propto \mathrm{sinc}(B t) * g_\text{SSMF}(t)$ and $g_\text{SSMF}(t)$ is the linear response of a standard single-mode fiber (SSMF). The SLD outputs $\nonlp{\cdot} = \lvert \,\cdot\, \rvert^2$ and the front-end of the ADC is a brickwall filter $h(t) \propto \mathrm{sinc}(2 B t)$ with twice the transmit filter bandwidth. 

\subsection{Discrete Time Model}
\label{sec:time-discrete_model}
We collect samples $Y_k = Y( k T_\text{s}')$, $k\in\mathbb{Z}$, where $T_\text{s}' = 1/(B N_\text{os})$ corresponds to sampling at rate $B N_\text{os}$ with oversampling factor $N_\text{os}$.
The Nyquist-Shannon sampling theorem is met if $Y(t)$ is bandlimited and $N_\text{os}$ is sufficiently high. The $k^\text{th}$ receiver sample may be written as
\begin{equation}
\begin{aligned}
    Y_k =  Z_k + N_k
     \label{eq:y_time_discrete}
\end{aligned}
\end{equation}
where
\begin{align*}
    Z_k 
    = \left[ h(t) * \nonlp{X(t)}
    \right]_{t=kT_\text{s}'}, \quad
    N_k  = [ h(t) * N'(t)]_{t = k T_\text{s}'} .
\end{align*}
The discrete-time noise $N_k$ is stationary, circularly-symmetric, and complex Gaussian with ACF $\varphi_\text{NN}[k] = N_0 \!\cdot\! \,\varphi_\text{NN}(k T_\text{s}')$.

We address the bandwidth expansion via oversampled simulations. Let $T_\text{sim} = T_\text{s}/N_\text{sim}$ be the simulation sampling period where $N_\text{sim}$ is the simulation oversampling factor and $d = N_\text{sim}/N_\text{os}$ is a positive integer. One can approximate
\begin{align}
   Z_k
   \approx \sum\nolimits_{\osidx'} h_{\osidx'}  \nonl\!\left(\sum\nolimits_{\osidx} g_{\osidx}  X'_{\left(d\cdot k-\osidx'\right)-\osidx}\right)
   \label{eq:disc_approx_z_k}
\end{align}
where $(X'_\osidx)_{\osidx\in\mathbb{Z}}=((0,\ldots,0,X_\kappa))_{\kappa \in\mathbb{Z}}$ is a $N_\text{sim}$-fold upsampled string, and $g_\osidx = g(\osidx T_\text{sim})$, $h_\osidx = h(\osidx T_\text{sim})$ are the oversampled filters. 
The SLD example above has $d_{\nonl} = 2$ and $N_\text{os}=N_\text{sim}=2$ results in sufficient statistics, and~\eqref{eq:disc_approx_z_k} is an equality. However, if $\nonlp{\cdot}$ is not a polynomial, $N_\text{sim}$ must be chosen sufficiently large so that~\eqref{eq:disc_approx_z_k} is a good approximation.

To illustrate, assume filters $g_\osidx$ and $h_\osidx$ in~\eqref{eq:disc_approx_z_k} with odd lengths $K_g$ and $K_h$, respectively. Suppose 
their taps are zero outside the intervals
$[-\lfloor K_g/2\rfloor,\lfloor K_g/2\rfloor]$ and $[-\lfloor K_h/2\rfloor,\lfloor K_h/2\rfloor]$.
The filters have symbol memories $\widetilde{K}_g = \lfloor (K_g-1)/N_\text{sim} \rfloor$ and $\widetilde{K}_h = \lfloor (K_h-1)/N_\text{sim} \rfloor$, and the total system memory is 
\begin{align}
    \widetilde{K} = \widetilde{K}_g + \widetilde{K}_h
    .  
    \label{eq:total_memory}
\end{align}

We transmit blocks with $\lfloor K_g/2\rfloor + \lfloor K_h/2\rfloor$ zeros at the beginning and end of each block. We collect $n$ transmit symbols in the vector $\mathbf{x}$ and the corresponding $N_\text{os}$ channel outputs per transmitted symbol in the vector $\mathbf{y}$, respectively: 
\begin{align*}
    \mathbf{x} = [x_1, \dots,\,  x_{n} ]^\mathrm{T}  \dimC{n}
    \;,\;\;\;
    \mathbf{y} = [y_1, \dots,\,  y_{N_\text{os}  n} ]^\mathrm{T}  \dimC{N_\text{os} n}.
\end{align*}

\section{SDD and SIC Rates}
\label{sec:detection_and_decoding}
\subsection{SDD Rates}
SDD passes symbol-wise APPs $P_{X_\kappa | \mathbf{Y}}(\cdot \rvert \mathbf{y})$, $\kappa \in \lbrace 1, \ldots, n \rbrace$ to the decoder for data estimation~\cite[Sec. III]{prinz2023successive}. 
A lower bound on the information rate of $\mathbf{X}$ and $\mathbf{Y}$:
\begin{align}
    I_n(\bm{X};\bm{Y}) %
    \ge \frac{1}{n} \left( H(\bm{X}) \!-\! \sum\nolimits_{\kappa=1}^n \! H(X_\kappa|\bm{Y})\right) := I_{n,\text{SDD}} \label{eq:separate_detection_decoding_rate} 
\end{align}
with equality if and only if $X_\kappa - \mathbf{Y} - X^{\kappa-1}$. Define the limits
\begin{align}
    I(\mathcal{X};\mathcal{Y}) := \lim_{n \rightarrow \infty} I_n(\bm{X};\bm{Y}), \quad
     I_\text{SDD} := \lim_{n\rightarrow \infty} I_{n,\text{SDD}}
\end{align}
so~\eqref{eq:separate_detection_decoding_rate} gives $I_\text{SDD} \le I(\mathcal{X};\mathcal{Y})$.
JDD achieves the rate $I(\mathcal{X};\mathcal{Y})$, but is usually too complex~\cite{huber1992trelliscodierte,schuh_reduced_2013,giallorenzi1996multiuser}. In practice, one is often limited to SDD, but~\cite[Fig.~7-9]{muller_capacity_separate2004} and~\cite[Fig.~6]{prinz2023successive} show that SDD experiences large rate losses in systems with memory. 

\subsection{SIC Rates}
\label{sec:sic}
We use SIC with $S$ stages and a different forward error control (FEC) code for each stage. To encode, downsample $\bm{X}$ by a factor of $S$ to create $S$ sequences of length $\siclength = n/S$; assume $\siclength$ is an integer. The symbols in the $s^\text{th}$ SIC stage are
\begin{align}
    \bm{V}_{s} &= (V_{s,t})_{t=1}^N = \big(X_{\kappa(s,1)}, X_{\kappa(s,2)} \ldots X_{\kappa(s,N)}   \big)
    \label{eq:subsampling}
\end{align}
where $\kappa(s,t) = s + (t-1)S$ converts a parallel indexing $(s,t)$ to a serial indexing $\kappa(s,t)$. See Fig.~\ref{fig:SP_conversion} as an example with $S=3$, $n=15$ and $N=5$ where $\bm{V}_{1} = \left(X_{1}, X_{4}, X_{7}, X_{10}, X_{13} \right).$

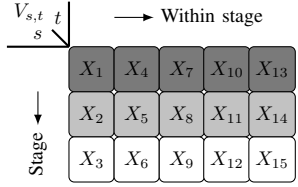
\begin{figure}
    \centering
    \usetikzlibrary{decorations.markings}
\tikzset{node distance=0.8cm}

\pgfdeclarelayer{background}
\pgfdeclarelayer{foreground}
\pgfsetlayers{background,main,foreground}

\pgfmathsetmacro{\samplerwidth}{30}

\tikzset{redbox/.style = {rounded corners=2pt,draw,font=\normalsize,minimum height=0.8cm,minimum width=0.8cm,fill=mycolorlightorange,}}
\tikzset{bluebox/.style = {rounded corners=2pt,draw,font=\normalsize,minimum height=0.8cm,minimum width=0.8cm,fill=mycolorcube,}}
\tikzset{bbox/.style = {rounded corners=2pt,draw,font=\normalsize,minimum height=0.8cm,minimum width=0.8cm,fill=mycolor5bright,}}
\tikzset{dot/.style = {anchor=base,fill,circle,inner sep=1pt}}

\begin{tikzpicture}[scale=0.75, transform shape]
\renewcommand{\baselinestretch}{1}

\node[redbox] (u1) {$X_1$};
\node[bluebox,below of=u1] (u2) {$X_2$};
\node[bbox,below of=u2] (u3) {$X_3$};

\node[redbox,right of=u1] (uM1) {$X_4$};
\node[bluebox,below of=uM1] (uM2) {$X_5$};
\node[bbox,below of=uM2] (uM3) {$X_6$};

\node[right of=uM1,redbox] (u2M1) {$X_7$};
\node[below of=u2M1,bluebox] (u2M2) {$X_{8}$};
\node[bbox,below of=u2M2] (u2M3) {$X_{9}$};

\node[redbox,right of=u2M1,minimum width=0.8cm] (u3M1) {$X_{10}$};
\node[bluebox,below of=u3M1,minimum width=0.8cm] (u3M2) {$X_{11}$};
\node[bbox,below of=u3M2] (u3M3) {$X_{12}$};

\node[redbox,right of=u3M1,minimum width=0.8cm] (u4M1) {$X_{13}$};
\node[bluebox,below of=u4M1,minimum width=0.8cm] (u4M2) {$X_{14}$};
\node[bbox,below of=u4M2] (u4M3) {$X_{15}$};

\draw[thick ] (u1) -- (-0.7,+0.7) node[above left]{$V_{s,\sicindex}$};
\draw[thick ] ($(u1) + (-0.4,+0.4)$) -- node[above](m){$s$} (-1.5,+0.4);
\draw[thick ] ($(u1) + (-0.4,+0.4)$) -- node[left,yshift=0.12cm](r){$\sicindex$} (-0.4,+1.2);

\draw[-latex] ($(m) + (0,-1.0)$) -- ($(m) + (0,-1.6)$)  node[left,rotate=90]{Stage}; 
\draw[-latex] ($(r) + (+1.0,0)$) -- ($(r) + (1.7,0)$) node[right]{Within stage};

\end{tikzpicture}
    \vspace{-6pt}
    \caption{Encoding with $S=3$ stages, $N=5$ and $n=15$ input symbols.}
    \label{fig:SP_conversion}
\end{figure}
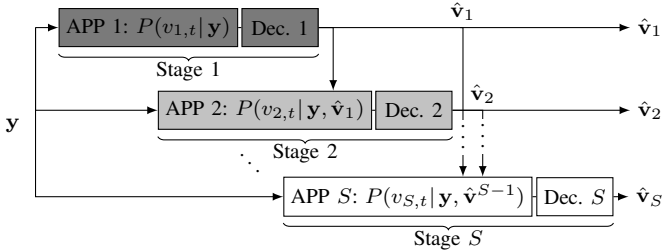
\begin{figure}
    \centering
    \vspace{-1pt}
    \usetikzlibrary{decorations.markings}
\tikzset{node distance=1cm}

\pgfdeclarelayer{background}
\pgfdeclarelayer{foreground}
\pgfsetlayers{background,main,foreground}

\pgfmathsetmacro{\samplerwidth}{30}

\tikzset{redbox/.style = {draw,minimum height=1.8em,minimum width=3.6em,fill=mycolorlightorange}}
\tikzset{bluebox/.style = {draw,minimum height=1.8em,minimum width=3.6em,fill=mycolorcube}}
\tikzset{greenbox/.style = {draw,minimum height=1.8em,minimum width=3.6em,, opacity=1,fill=mycolor5bright},
mydash/.style={rectangle,fill=white}}
\tikzset{dot/.style = {anchor=base,fill,circle,inner sep=1pt}}
\tikzstyle{point}=[fill,shape=circle,minimum size=3pt,inner sep=0pt]

\begin{tikzpicture}[]
\footnotesize

\node[] (y) {};
\node[redbox,right of=y,align=center,node distance=1.5cm,font=\footnotesize,label={[xshift=+0.5cm,yshift=-0.05cm]above:{}}] (app1) {APP 1: $P(v_{1,t} |\, \mathbf{y})$};
\node[redbox,right of=app1,align=center,node distance=1.75cm] (dec1) {Dec. 1};

\node[bluebox,below of=app1,yshift=-0.10cm,xshift=1.55cm,align=center,label={[xshift=0.72cm,yshift=-0.05cm]above left:{}}] (app2) {APP 2: $P( v_{2,t} |\, \mathbf{y}, \hat{\mathbf{v}}_1)$};
\node[bluebox,right of=app2,node distance=1.98cm,align=center] (dec2) {Dec. 2};

\node[below of=app2,node distance=0.70cm,xshift=-0.2cm,yshift=0.1cm] (recDots) {$\ddots$};
\node[greenbox,below of=app2,yshift=-0.15cm,xshift=1.9cm,align=center,label={[xshift=0.7cm,yshift=-0.05cm]above left:{}}] (appM) {APP $S$: $P( v_{S,t} |\, \mathbf{y}, \hat{\mathbf{v}}^{S-1})$};
\node[greenbox,right of=appM,node distance=2.22cm,align=center] (decM) {Dec. $S$};

\draw[-latex] (app1 -|  y) -- (app1);
\draw (app1) --(dec1);

\draw[-latex] (app1 -| y) |- (appM);
\draw[-latex] (app1 -| y) |- (app2);

\node[xshift=-0.3cm,yshift=-0.2cm](y_desc) at(y |- app2) {$\mathbf{y}$}; 

\draw[-latex] (dec1) -| node[mydash,pos=0.87]{$\rvdots$} (appM.20) node[midway,above](v1label) {$\hat{\mathbf{v}}_1$};
\draw[-latex] (dec1) -| (app2.16);
\draw[-latex] (app2) --  (dec2) -| node[mydash,pos=0.70]{$\rvdots$} (appM.15) node[midway,above] {$\hat{\mathbf{v}}_2$};

\node[xshift=1.0cm,font=\footnotesize](out1) at(dec1 -| decM) {$\hat{\mathbf{v}}_1$}; 
\node[xshift=1.0cm,font=\footnotesize](out2) at(dec2 -| decM) {$\hat{\mathbf{v}}_2$};
\node[xshift=1.0cm,font=\footnotesize](outM) at(decM -| decM) {$\hat{\mathbf{v}}_S$};

\draw (appM) -- (decM); 
\path[] (app1) -- (appM);

\draw[-latex,] (dec1 -| appM.21) -- (out1);
\draw[-latex,] (dec2) -- (out2);
\draw[-latex,] (decM) -- (outM);

\draw [decorate, decoration = {mirror,brace}] ($(app1.west) + (0,-0.35)$) -- node[midway,below,font=\footnotesize] {Stage 1} ($(dec1.east) + (0,-0.35)$);
\draw [decorate, decoration = {mirror,brace}] ($(app2.west) + (0,-0.35)$) -- node[midway,below,font=\footnotesize] {Stage 2} ($(dec2.east) + (0,-0.35)$);
\draw [decorate, decoration = {mirror,brace}] ($(appM.west) + (0,-0.35)$) -- node[midway,below,font=\footnotesize] {Stage $S$} ($(decM.east) + (0,-0.35)$);

\end{tikzpicture}
    \vspace*{-22pt}
    \caption{SIC receiver with SDD for each stage.}
    \label{fig:sic}
\end{figure}

We detect and decode in $S$ stages; see Fig.~\ref{fig:sic}. To illustrate, consider again $S=3$ and partition $\bm{X}$ into $\mathbf{V}_1$, $\mathbf{V}_2$, $\mathbf{V}_3$. 
The first stage performs SDD and calculates APPs for $\mathbf{V}_1$. %
A decoder estimates $\hat{\mathbf{V}}_1$.
The second stage uses $\hat{\mathbf{V}}_1$ as prior information and calculates conditional APPs for 
$\mathbf{V}_2$.
Correct prior information increases the information rate in the second stage. The final stage calculates APPs for $\mathbf{V}_3$, given $(\hat{\mathbf{V}}_1, \hat{\mathbf{V}}_2)$. 

Collect $\bm{V} = (\bm{V}_s)_{s=1}^S$. Since $\mathbf{V}$ is a reordered $\mathbf{X}$, we have $I_n(\bm{X};\bm{Y}) =I_n(\bm{V};\bm{Y})$. 
The average SIC rate is~\cite[Sec. IV]{prinz2023successive}
\begin{align}
    I_\text{$n$,SIC} = \frac{1}{S} \sum\nolimits_{s=1}^S I^{s}_{N\!,\text{SIC}}
\end{align}
with stage rates $I^{s}_{N\!,\text{SIC}} := 1/\siclength  \sum_{t = 1}^N I(V_{s,t}; \bm{Y},\bm{V}^{s-1})$ and inequality $I_\text{$n$,SDD} \le I_\text{$n$,SIC} \le  I_n(\bm{V};\bm{Y})$. The limits  are $I^{s}_{\text{SIC}} := \lim_{N \rightarrow \infty } I^{s}_{N\!,\text{SIC}}$ and $I_{\text{SIC}} := \lim_{n \rightarrow \infty } I_{n,\text{SIC}}$. We encode $\mathbf{V}_s$ with a code rate less than $I^{s}_{\text{SIC}}$ to ensure reliable decoding as the block length grows, i.e., we may assume  $\hat{\mathbf{V}}_s = \mathbf{V}_s$; see~\cite{PfisterAIRFiniteStateChan2001}. 
Comparing the limiting rates, we obtain
\begin{align}
     I_\text{SDD} \le I_\text{SIC} \le I(\mathcal{X};\mathcal{Y}) .
    \label{eq:sdd_jdd_inequality}
\end{align}
SIC can approach the JDD performance by increasing $S$~\cite[Fig.~6]{prinz2023successive}. Note that SDD has $S=1$.

\section{NN-APP Detector}
\label{sec:app_nn}
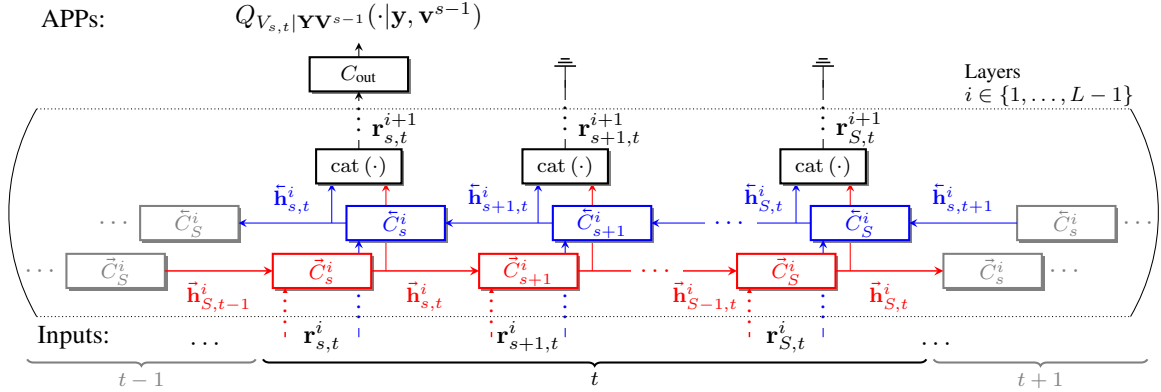
\begin{figure*}[!ht]
    \centering
    \hspace*{-25pt}
    \resizebox{1.85\columnwidth}{!}
    {
    \tikzset{
block/.style = {drop shadow={shadow xshift=0.1em,shadow yshift=-0.1em,}, draw, thick, fill=white, fill opacity=1,rectangle, minimum height=1.32em, minimum width=3.8em,font=\footnotesize,outer sep=0,inner sep=0},
conblock/.style = {black,draw=black,drop shadow={shadow xshift=0.1em,shadow yshift=-0.1em,}, draw, thick, fill=white, fill opacity=1,rectangle, minimum height=1.32em, minimum width=3.3em,font=\footnotesize,outer sep=0,inner sep=0},
sum/.style= {draw, fill=white, circle, node distance=1cm,inner sep=0,outer sep=0},
input/.style = {coordinate},
output/.style= {coordinate},
pinstyle/.style = {pin edge={to-,thin,black}},
mydash/.style={anchor=center,fill=white,rectangle},
graystyle/.style={draw=gray,opacity=1,text=gray}
}

\pgfdeclarelayer{background}
\pgfdeclarelayer{foreground}
\pgfsetlayers{background,main,foreground}   %

\begin{tikzpicture}[auto, red, draw=red, node distance=3.0cm,>=stealth,line width=0.2mm]

    \node[input](input){}; 
    \node[block,right of=input,node distance=2cm](C1){$\Cfw{s}{i}$}; 
    \node[block,right of=C1,node distance=2.8cm](C2){$\Cfw{s+1}{i}$}; 
    \node[block,right of=C2,node distance=3.5cm](C3){$\Cfw{S}{i}$}; 
    
    \node[output,below of=C1,node distance=0.9cm](out1){}; 
    \node[output,below of=C2,node distance=0.9cm](out2){}; 
    \node[output,below of=C3,node distance=0.9cm](out3){}; 
    
    \def\stackshift{0.5cm} 
    \node[conblock,above of=C1,node distance=0.9cm,xshift=\stackshift,yshift=0.52cm](conblock1){$\vstack{(\cdot)}$}; 
    \node[conblock,above of=C2,node distance=0.9cm,xshift=\stackshift,yshift=0.52cm](conblock2){$\vstack{(\cdot)}$}; 
    \node[conblock,above of=C3,node distance=0.9cm,xshift=\stackshift,yshift=0.52cm](conblock3){$\vstack{(\cdot)}$}; 
    
    \node[output,above of=conblock1,node distance=0.7cm](output1){}; 
    \node[output,above of=conblock2,node distance=0.7cm](output2){}; 
    \node[output,above of=conblock3,node distance=0.7cm](output3){}; 
    
    \node[draw=black,black,block,above of=output1,node distance=0.55cm](app1){$C_\text{out}$}; 
    \node[black,above of=output2,node distance=0.6cm](app2){}; 
    \node[black,above of=output3,node distance=0.6cm](app3){}; 
    
    \node[block,left of=C1,node distance=2.8cm,graystyle](C0){$\Cfw{S}{i}$}; 
    \node[block,right of=C3,node distance=2.8cm,graystyle](C4){$\Cfw{s}{i}$}; 
    
    \node[left of=C0,node distance=1.0cm,graystyle,draw=none](){$\cdots$}; 
    \node[right of=C4,node distance=1.0cm,graystyle,draw=none](){$\cdots$}; 
    
    \draw[->] (out1)node[black]{$\mathbf{r}^i_{s,t}$}  
    ++(-\stackshift,0) node(inter1){} --node[mydash]{$\rvdots$} (C1.south -| inter1); 
    \draw[->] (out2)node[black]{$\mathbf{r}^i_{s+1,t}$} ++(-\stackshift,0) node(inter2){} --node[mydash]{$\rvdots$} (C2.south -| inter2);
    \draw[->](out3)node[black]{$\mathbf{r}^i_{S,t}$} ++(-\stackshift,0) node(inter3){} --node[mydash]{$\rvdots$} (C3.south -| inter3);
    
    \draw[->] (C0) -- node[below,font=\footnotesize]{$\hfw{S,t-1}{i}$} (C1);
    \draw[->] (C1) -- node[below,font=\footnotesize]{$\hfw{s,t}{i}$} (C2);
    \draw[->] (C2) --node[mydash,pos=.5] {$\dots$}  (C3) node[below ,xshift=-1.1cm,font=\footnotesize]{$\hfw{S-1,t}{i}$}; 
    \draw[->] (C3) --node[below,font=\footnotesize]{$\hfw{S,t}{i}$} (C4);
    
    \begin{pgfonlayer}{background}
        \draw[red,->] (C1.east) ++(0.2,0) node[](inter1){} -- (conblock1.south -| inter1);
        \draw[red,->] (C2.east) ++(0.2,0) node[](inter2){} -- (conblock2.south -| inter2);
        \draw[red,->] (C3.east) ++(0.2,0) node[](inter3){} -- (conblock3.south -| inter3);
    \end{pgfonlayer}

    \draw[black,->] (conblock1) node[above right,xshift=0.05cm,yshift=0.15cm]{$\mathbf{r}^{i+1}_{s,t}$} -- node[mydash,pos=0.5]{$\rvdots$} (app1); 
    
    \draw[black,-] (conblock2) node[above right,xshift=0.05cm,yshift=0.15cm]{$\mathbf{r}^{i+1}_{s+1,t}$} --node[mydash,pos=.425]{$\rvdots$} (app2); 
    \draw[black,-] (conblock3) node[above right,xshift=0.05cm,yshift=0.15cm]{$\mathbf{r}^{i+1}_{S,t}$} --node[mydash,pos=.425]{$\rvdots$} (app3);
    
    \draw[->,black] (app1) -- ++(0,0.43) node[above](appdesc){$Q_{V_{s,t}|\mathbf{Y}\mathbf{V}^{s-1}}(\cdot|\mathbf{y},\mathbf{v}^{s-1})$};
    
    \node[black,ground,line width=0.3pt,rotate around={180:(app2)},scale=0.7,yshift=0.15cm] at (app2.north){};
    \node[black,ground,line width=0.3pt,rotate around={180:(app3)},scale=0.7,yshift=0.15cm] at (app3.north){};

    \node[below left,text=black,yshift=0.1cm,xshift=-1.2cm] at(out1){$\cdots$};
    \node[below right,text=black,yshift=0.1cm,xshift=+1.7cm] at(out3){$\cdots$};

    \begin{pgfonlayer}{background}
    \begin{scope}[draw=blue,blue,yshift=0.65cm,xshift=2*\stackshift]
        \node[input](input){}; 
        \node[block,right of=input,node distance=2cm](C1){$\Cbw{s}{i}$}; 
        \node[block,right of=C1,node distance=2.8cm](C2){$\Cbw{s+1}{i}$}; 
        \node[block,right of=C2,node distance=3.5cm](C3){$\Cbw{S}{i}$}; 
        
        \node[block,left of=C1,node distance=2.8cm,graystyle](C0){$\Cbw{S}{i}$}; 
        \node[block,right of=C3,node distance=2.8cm,graystyle](C4){$\Cbw{s}{i}$}; 
        
        \node[left of=C0,node distance=1.0cm,graystyle,draw=none](){$\cdots$}; 
        \node[right of=C4,node distance=1.0cm,graystyle,draw=none](){$\cdots$};

        \draw[<-] (C0) --node[above,font=\footnotesize]{$\hbw{s,t}{i}$} (C1);
        \draw[<-] (C1) --  node[above,font=\footnotesize]{$\hbw{s+1,t}{i}$}
        (C2);
        \draw[<-] (C2) --node[mydash,pos=.5] {$\dots$} (C3) node[above ,xshift=-1.3cm,font=\footnotesize]{$\hbw{S,t}{i}$} ; 
        \draw[->] (C1.west) ++(-0.2,0) node[](inter1){} -- (conblock1.south -| inter1);
        \draw[->] (C2.west) ++(-0.2,0) node[](inter2){} -- (conblock2.south -| inter2);
        \draw[->] (C3.west) ++(-0.2,0) node[](inter3){} -- (conblock3.south -| inter3);

        \draw[<-] (C3) --node[above,font=\footnotesize] {$\hbw{s,t+1}{i}$} (C4);
        
        \draw[->] (out1) ++(\stackshift,0) node(inter1){} --node[pos=.3,mydash]{$\rvdots$} (C1.south -| inter1); 
        \draw[->] (out2) ++(\stackshift,0) node(inter2){} --node[pos=.3,mydash]{$\rvdots$} (C2.south -| inter2); 
        \draw[->] (out3) ++(\stackshift,0) node(inter3){} --node[pos=.3,mydash]{$\rvdots$} (C3.south -| inter3); 
    \end{scope}
    \end{pgfonlayer}

\draw [
    black,
    thick,
    decoration={
        brace,
        mirror,
        raise=0.5cm
    },
    decorate,
    font=\footnotesize
] ($ (out1) + (-0.8,+0.2) $) -- ($ (out3) + (1.9,+0.2) $)
node [pos=0.5,anchor=north,yshift=-0.55cm] {$t$}; 

\draw [
    gray,
    thick,
    decoration={
        brace,
        mirror,
        raise=0.5cm
    },
    decorate,
    font=\footnotesize
] ($ (out1) + (8.3,+0.2) $) -- ($ (out3) + (4.9,+0.2)$) node [pos=0.5,anchor=north,yshift=-0.55cm] {$t+1$}; 

\draw [
    gray,
    thick,
    decoration={
        brace,
        mirror,
        raise=0.5cm
    },
    decorate,
    font=\footnotesize
] ($ (out1) + (-4.0,+0.2) $) -- ($ (out3) + (-7.2,+0.2) $)
node [pos=0.5,anchor=north,yshift=-0.55cm] {$t-1$};

\draw [black] ($(C0 -| out1) + (-1.1,-0.05)$) ++(150:3.2) node(arclstart){} arc (150:210:2.82) node(arclend){};
\draw [black] ($(C0 -| out1) + (8.2,-0.05)$) ++(30:3.2) node(arcrstart){}  arc (30:-30:2.82) node(arcrend){};
\node[black,yshift=-0.20cm,xshift=-1cm,text width=2.5cm,font=\footnotesize] at (arcrstart |- app3){Layers \\ $i \in \{1,\ldots,\maxlay-1\}$};

\draw[black,densely dotted] (arclstart.center) -- (arcrstart.center); 
\draw[black,densely dotted] (arclend.center) -- (arcrend.center); 
   
\node [black,xshift=0.5cm] at(arclstart |- out3){Inputs:};
\node [black,xshift=0.5cm] at(arclstart |- appdesc){APPs:};

\end{tikzpicture}
    }
    \vspace{-14pt}
    \caption{Bidirectional time-varying RNN for SIC stage $s$.}
    \label{fig:tvrnn_architecture}
\end{figure*}

Running the FBA in every SIC stage is too complex even for relatively small memory $\widetilde{K}$ or alphabets $\mathcal{A}$, so one must use mismatched models. We propose an NN with an FBA structure that is recurrent and periodically time-varying~\cite{plabst2024neural}. Fig.~\ref{fig:tvrnn_architecture} shows the recurrent NN (RNN), which consists of multiple layers. Each layer is bidirectional with a forward and backward path and internal states similar to the FBA~\cite{kim2020physical,haykin2000adaptive}. 

\subsection{NN Structure and Processing}
Suppose the NN inputs are real-valued or use composite-real representations for complex-valued inputs. In SIC stage $s$ we process inputs for the remaining $j = s,\ldots,S$ stages:
\begin{align}
    \mathbf{r}^1_{j,t} := (\overline{\mathbf{y}}_{j,t},\overline{\mathbf{v}}_{j,t}) \;\dimR{L_\mathrm{Y} + L_\mathrm{IC}}
    \label{eq:r_vec}
\end{align}
with $L_\mathrm{Y}$ channel outputs $\overline{\mathbf{y}}_{j,t}$ 
and $L_\mathrm{IC}$ symbols $\overline{\mathbf{v}}_{j,t}$. We use
\begin{align}
    \overline{\mathbf{y}}_{j,t} := ( y_{N_\text{os}  \cdot \kappa(j,t) +\, u } )_{u=-\Delta}^{\nabla} \;\dimR{L_\mathrm{Y}}
    \label{eq:chunk_obs}
\end{align}
where $\Delta:=\lfloor (L_\mathrm{Y}-1)/2\rfloor$ and $\nabla:=\lceil (L_\mathrm{Y}-1)/2\rceil$ correspond to symbols before and after transmission of the symbol $v_{j,t} = x_{\kappa(j,t)}$, respectively.
Next, collect the $L_\text{IC}$ symbols among $\mathbf{v}^{s-1}$ that are closest to $v_{j,t} = x_{\kappa(j,t)}$ in the vector
\begin{align}
   \overline{\mathbf{v}}_{j,t} :=  ( x_\kappa \mid \kappa \in  \mathcal{V}_{j,t} ) \; \dimR{L_\text{IC}} 
   \label{eq:chunk_ic}
\end{align}
where ``closest" means
\begin{align}
    \mathcal{V}_{j,t} = \argmin_{\lvert \mathcal{U}\rvert = L_\text{IC}}   \; \sum\nolimits_{\substack{a \in \mathcal{U},\, x_a \in \mathbf{v}^{s-1}}}\; |a - \kappa(j,t)| .
    \label{eq:set_tkappa}
\end{align}
We use zero-padding to extend the vectors~\eqref{eq:chunk_obs} where needed.

RNN inputs are processed in forward and backward order:
\begin{align}
\begin{array}{llllll}
    \ldots & \mathbf{r}^1_{s,t}, & \mathbf{r}^1_{s+1,t}, & \ldots, & \mathbf{r}^1_{S,t}, & \\
    & \mathbf{r}^1_{s,t+1},& \mathbf{r}^1_{s+1,t+1},& \ldots,& \mathbf{r}^1_{S,t+1}& \ldots
    \end{array}
    .
    \label{eq:unrolled_input}
\end{align}
and stage $s$ processes a total of $N (S-s+1)$ inputs. The input-process~\eqref{eq:unrolled_input} for stages $s$ to $S$ is cyclostationary with period $S\!-\!s\!+\!1$ due to the SIC partitioning of Sec.~\ref{sec:sic}. Likewise, the FBA state metrics are cyclostationary~\cite{plabst2024neural} with the same period. We thus adapt classic RNNs~\cite[Chap.~10.3]{goodfellow2016deep} to periodically time-varying input processing and state recursions.

Fig.~\ref{fig:tvrnn_architecture} shows the structure of such a time-varying RNN with $\maxlay$ layers: $\maxlay-1$ layers are recurrent with forward and backward paths, and the last layer is feedforward. 
Consider the forward path in layer $i$. This path has cells $\Cfw{s}{i}, \Cfw{s+1}{i}, \ldots, \Cfw{S}{i}$ that repeat periodically for $t=1,\ldots,N$. Compared to classic RNNs~\cite[Chap.~10.3]{goodfellow2016deep}, the number of RNN parameters increases by a factor of $S-s+1$, but the computational complexity remains the same. 
RNNs have internal states $(\fw{\mathbf{h}}^i_{j,t})_{j=s}^S$ in the forward and backward paths, where $t= 1,\ldots, N$.

To illustrate further, let the input dimensions of the recurrent layers and the output layer be $(\lay_1, \ldots, \lay_{\maxlay-1}, \lay_\maxlay)$, where $\lay_1 = L_\mathrm{Y} + L_\text{IC}$. We convert the pairs~\eqref{eq:r_vec} to column vectors. The state recursion via cells $\Cfw{j}{i}$, $j = s,\ldots,S$, is
\begin{align}
    \hfw{j,t}{i} = f\big(
    \Wfw{\text{in},{j}}{i}( \mathbf{r}_{j,t}^i)  + \Wfw{j-1}{i}( \hfw{j-1,t}{i} )\big)
    \; \dimR{\lay_{i+1}/2}
\end{align}
where $f(\cdot)$ is the element-wise rectified linear unit (ReLU), and the index pair $(s-1,t) \mapsto (S,t-1)$ as well as the single index $s-1 \mapsto S$ due to the parallel indexing.
The input and state transformations for $j = s,\ldots,S$ are: 
\begin{align*}
    &\Wfw{\text{in},j}{i}: \mathbb{R}^{\lay_{i}} \mapsto \mathbb{R}^{\lay_{i+1}/2}, && 
    \Wfw{\text{in},j}{i}(\mathbf{r}_{j,t}^i) = \fw{\mathbf{W}}_{\text{in},j}^i \mathbf{r}_{j,t}^i +  \fw{\mathbf{b}}^i_{\text{in},j}
    \\
    &\Wfw{j}{i}: \mathbb{R}^{\lay_{i+1}/2} \mapsto \mathbb{R}^{\lay_{i+1}/2}, && \Wfw{j}{i}(\hfw{j,t}{i}) = \fw{\mathbf{W}}^i_j \hfw{j,t}{i} +  \fw{\mathbf{b}}^i_j
\end{align*}
where $\fw{\mathbf{W}}_{\text{in},j}^i$, $\fw{\mathbf{W}}_j^i$ and $\fw{\mathbf{b}}_{\text{in},j}^i$, $\fw{\mathbf{b}}^i_j$ are input and state recursion matrices, and input and state bias vectors, respectively. The backward path works analogously; see Fig.~\ref{fig:tvrnn_architecture}. 

To complete the recurrent layer processing, the outputs of the two paths are concatenated for $j = s,\ldots,S$ and all $t$:  
\begin{align}
    \mathbf{r}_{j,t}^{i+1} = \vstack( \hfw{j,t}{i}, \hbw{j,t}{i}) \;\;\dimR{\lay_{i+1}}
    \label{eq:rnn_output_concat}
\end{align}
and \eqref{eq:rnn_output_concat} is passed to the next recurrent layer if $i+1 < \maxlay$. If $i + 1 = \maxlay$, the last cell $C_\text{out}$ performs final processing, i.e., for all $t$ and fixing $j=s$ we have
\begin{align}
    Q_{V_{s,t}|\mathbf{Y},\mathbf{V}^{s-1}}(\cdot|\mathbf{y},\mathbf{v}^{s-1}) = 
    \phi( 
    \mathbf{W}_\text{out} \mathbf{r}_{s,t}^{\maxlay} + \mathbf{b}_\text{out}) \;\dimR{\lvert \mathcal{A} \rvert}
    \label{eq:app_sic_s_rnn}
\end{align}
where $\mathbf{W}_\text{out} \dimR{|\mathcal{A}| \times \lay_\maxlay}$, $\mathbf{b}_\text{out} \dimR{|\mathcal{A}|}$,
and the ``softmax'' function $\phi(\cdot)$~\cite{bishop2006pattern} generates a PMF interpreted as symbol-wise APPs. To initialize the RNN, we set the first forward and last backward states in all recurrent layers to zero.

\subsection{Achievable Rates and NN Optimization}
\label{sec:ar_and_nn_optim}
The NN-APP detector approximates APPs~\eqref{eq:app_sic_s_rnn}. A lower-bound on the SIC rate is (see~\cite[Sec.~VI~D.]{prinz2023successive},~\cite{arnoldsimulationmi})
\begin{align*}
    I_{n,\text{SIC}} \geq
    \frac{1}{S}  \!\sum_{s=1}^S \! \underbrace{\frac{1}{\siclength}  \sum_{t=1}^N \! H(V_{s,t}) + \mathbb{E}[\log_2 Q(V_{s,t} | \bm{Y},\!\bm{V}^{s-1})]}_{\textstyle := I^{s}_{q,N\!,\text{SIC}}}
\end{align*}
where the expectation is over the actual $p_{V_{s,t}, \mathbf{Y}, \mathbf{V}^{s-1}}$. The expression $I^{s}_{q,N\!,\text{SIC}}$ is the  mismatched rate of SIC stage $s$ and we define the limiting rate as $I^{s}_{q,\text{SIC}} := \lim_{N\rightarrow \infty} I^{s}_{q,N\!,\text{SIC}}$.

We wish to maximize $I^{s}_{q,N\!,\text{SIC}}$ for each SIC stage $s$. %
We estimate $I^{s}_{q,N\!,\text{SIC}}$ via simulation and formulate the optimization problem as a cross-entropy minimization~\cite[Sec.~4.1]{bocherer2022mlcomm}
\begin{align}
    \hspace{-6pt}\argmin_{(\!Q_{V_{s,t} | \mathbf{Y}, \mathbf{V}^{s-1}}\!)_{t=1}^N} \!
    \hspace{-4pt} - 
    \frac{1}{\siclength} \! \sum_{t=1}^\siclength  
    \!\left\langle \log_2 Q_{V_{s,t}|\mathbf{Y}, \mathbf{V}^{s-1}}(v_{s,t} | \mathbf{y},  {\mathbf{v}^{s-1}}) 
    \! \right\rangle
    \label{eq:nn_optim}
\end{align}
where $\langle \cdot \rangle$ denotes Monte-Carlo averaging over $N_\text{blk}$ transmit symbols and pairs $(\mathbf{v}^{(w)},\,\mathbf{y}^{(w)})_{w=1}^{N_\text{blk}}$ with the APPs~\eqref{eq:app_sic_s_rnn}. %

We optimize one NN per SIC stage and SNR. The NNs are initialized with optimized parameters from a lower SNR, if available. We approximate~\eqref{eq:nn_optim} via ADAM~\cite{kingma2017adam} and  perform stochastic batch gradient descent with strings $(\mathbf{v}^{(w)}, \mathbf{y}^{(w)})_{w=1}^{N_\text{batch}}$ generated each from $N = T_\text{RNN}/(S-s+1)$ channel inputs; the parameter $T_\text{RNN}$ limits the number of sequential RNN inputs to avoid numerical instabilities~\cite{pascanu2013difficulty}. We take $N_\text{iter}$ gradient steps with step size $\beta_\text{lr}$. The number of inputs $T_\text{RNN}$ and size $L_\text{Y}$ indicate the maximum symbol memory that the RNN can capture and is  $\widetilde{N}_\text{RNN} \approx \lfloor L_\text{Y}/N_\text{os} \rfloor + (T_\text{RNN}-1)$.

\section{Numerical Results}
\label{sec:numerical}
We study detection for short-haul fiber-optic links with a SLD $\nonlp{\cdot} = |\cdot|^2$, which is a phase-retrieval problem~\cite{schniter2015compressive}; see~Sec.~\ref{sec:fiber_optic_application}. The program code is available at~\cite{nnsiccode2024plabst}. 

\begin{table}
\caption{Algorithmic complexity per APP estimate.}
\vspace{-8pt}
\centering
{\renewcommand{\arraystretch}{1.0}
\begin{tabular}{ll} 
\toprule
\textbf{Algorithm} & \textbf{Multiplications}  \\ \midrule
FBA~\cite{plabst2022achievable} & $\mathcal{O}(S \cdot \lvert \mathcal{A} \rvert^{\widetilde{K} + 1})$\\ Bit-wise GS~\cite{prinz2023successive} & $\mathcal{O}(S \cdot \widetilde{K}^2 \cdot m \cdot N_\text{iter} \cdot N_\text{par})$\\
NN & $\mathcal{O}\big(S \cdot \big(\sum_{i=1}^{\maxlay-1} \lay_{i}\lay_{i+1} + \lay_{i+1}^2/2  + \lay_\maxlay \cdot |\mathcal{A}|  \big)\big)$\\
\bottomrule
\end{tabular}}
\label{tab:complexity}
\end{table}

\begin{table*}[!htb]
\centering
\caption{NN Parameters.}
\vspace{-9pt}
\label{tab:nnparams}
\renewcommand{\arraystretch}{0.0} 

\pgfplotstableset{
    columns={modulation,specify,layer,nrnn,Ls,Tr,B,V,n,lr,I},
    font={\footnotesize},
    col sep = comma,
    every head row/.style={before row=\toprule,after row=\midrule},
    every last row/.style={after row=\bottomrule},
    columns/modulation/.append style={column type = {l},string type,column name={\textbf{Modulation}}},
    columns/specify/.append style={column type = {l},string type,column name={\textbf{}}},
    columns/layer/.append style={column type = {l},string type,string replace*={_}{, }, string replace*={, A}{}},
    columns/nrnn/.append style={column name={$\widetilde{N}_\text{RNN}$},string replace*={nrnn=}{}},
    columns/lr/.append style={column name={$\beta_\text{lr}$},string replace*={lr=}{}},
    columns/Tr/.append style={column name={$T_\text{RNN}$},string replace*={Tr=}{}},
    columns/B/.append style={column name={$N_\text{batch}$},string replace*={B=}{}},
    columns/S/.append style={column name={$S$},string replace*={S=}{}},
    columns/Ls/.append style={string replace*={Ls=}{},column name={$L_\text{IC}$}},
    columns/I/.append style={sci,column name={$N_\text{iter}$},string replace*={I=}{}},
    columns/V/.append style={sci,column name={$N_\text{blk}$},string replace*={V=}{}},
    columns/n/.append style={sci,column name={$n$},string replace*={n=}{}},
    columns/L/.append style={column name={$L_\text{fib}\,\mathrm{[km]}$},string replace*={L=}{},multiply by=1e-3},
    columns/Rs/.append style={column name={$B\,\mathrm{[GBd]}$},string replace*={Rs=}{},multiply by=1e-9},
    columns/C/.append style={sci,column name={$C_\text{mul}$},string replace*={C=}{},string replace*={.txt}{}},
}
\begin{filecontents*}{params_all.csv}
%% LaTeX2e file `params_all.csv'
%% generated by the `filecontents' environment
%% from source `isit2024' on 2024/05/21.
%%
ver,modulation,specify,layer,nrnn,lr,Tr,B,S,Ls,I,V,n,P,a,L,Rs,C
% v1.5,$M=4$,,32_64_A,nrnn=47,lr=1E-03,Tr=32,B=128,S=2,Ls=16,I=1.0E+04,V=1.0E+03,n=6.0E+04,P=151,a=0,L=0.0E+00,Rs=3.5E+10,C=5.4E+03.txt
% v1.5,$M=8$,,64_128_64_A,nrnn=95,lr=5E-04,Tr=64,B=64,S=2,Ls=32,I=6.0E+04,V=3.0E+03,n=6.0E+04,P=151,a=0,L=0.0E+00,Rs=3.5E+10,C=3.1E+04.txt
% v1.5,$M=16$,,84_128_128_A,nrnn=105,lr=2E-04,Tr=64,B=64,S=2,Ls=64,I=6.0E+04,V=3.0E+03,n=6.0E+04,P=151,a=0,L=0.0E+00,Rs=3.5E+10,C=5.4E+04.txt
% v1.5,$M=32$,,84_128_128_A,nrnn=105,lr=1E-04,Tr=64,B=64,S=2,Ls=64,I=6.0E+04,V=3.0E+03,n=6.0E+04,P=151,a=0,L=0.0E+00,Rs=3.5E+10,C=5.6E+04.txt
% v1.5,$M=64$,,84_128_128_128_64_A,nrnn=125,lr=5E-05,Tr=84,B=84,S=2,Ls=100,I=7.5E+04,V=3.0E+03,n=6.0E+04,P=151,a=0,L=0.0E+00,Rs=3.5E+10,C=9.5E+04.txt
% v1.5,$M=128$,,84_128_128_128_128_A,nrnn=125,lr=5E-05,Tr=84,B=84,S=2,Ls=100,I=7.5E+04,V=3.0E+03,n=6.0E+04,P=151,a=0,L=0.0E+00,Rs=3.5E+10,C=1.2E+05.txt
v1.5,$M=4$,,64_128_64_A,nrnn=95,lr=5E-04,Tr=64,B=128,S=2,Ls=32,I=2.0E+04,V=1.0E+03,n=6.0E+04,P=151,a=0,L=3.0E+04,Rs=3.5E+10,C=3.1E+04.txt%PAM/ASK
%v1.5,$M=4$,SQAM,64_256_128_128_A,nrnn=95,lr=5E-04,Tr=64,B=128,S=2,Ls=32,I=2.0E+04,V=1.0E+03,n=6.0E+04,P=151,a=0,L=3.0E+04,Rs=3.5E+10,C=1.3E+05.txt
v1.5,$M=8$,,64_128_128_A,nrnn=115,lr=3E-04,Tr=84,B=128,S=2,Ls=32,I=5.0E+04,V=1.0E+03,n=6.0E+04,P=151,a=0,L=3.0E+04,Rs=3.5E+10,C=4.6E+04.txt%PAM/ASK
%v1.5,$M=8$,SQAM,64_256_128_128_A,nrnn=115,lr=3E-04,Tr=84,B=128,S=2,Ls=32,I=5.0E+04,V=1.0E+03,n=6.0E+04,P=151,a=0,L=3.0E+04,Rs=3.5E+10,C=1.3E+05.txt
v1.5,$M=16$,,84_200_128_128_A,nrnn=161,lr=5E-05,Tr=120,B=64,S=6,Ls=64,I=8.0E+04,V=3.0E+03,n=8.0E+04,P=151,a=0,L=3.0E+04,Rs=3.5E+10,C=1.1E+05.txt
v1.5,$M=32$,,100_200_200_200_168_A,nrnn=169,lr=4E-05,Tr=120,B=64,S=6,Ls=64,I=1.0E+05,V=7.0E+03,n=8.0E+04,P=151,a=0,L=3.0E+04,Rs=3.5E+10,C=2.3E+05.txt
v1.5,$M=64$,,100_300_300_300_240_A,nrnn=169,lr=4E-05,Tr=120,B=64,S=6,Ls=100,I=1.0E+05,V=7.0E+03,n=8.0E+04,P=151,a=0,L=3.0E+04,Rs=3.5E+10,C=4.9E+05.txt
\end{filecontents*}
\pgfplotstabletypeset[
columns={modulation,specify,layer,nrnn,Ls,Tr,B,V,n,lr,I,L},
columns/layer/.append style={column name={$L_\mathrm{Y},\ell_2,\ell_3,\ell_4,\ell_5$}}]{params_all.csv}
\end{table*}

We transmit $N_\text{blk}$ blocks, each with $n \geq \SI{20e3}{}$ symbols, and evaluate the SIC rates~\eqref{eq:nn_optim} via Monte Carlo simulation. The DAC performs sinc pulse shaping at the symbol rate \SI{35}{\giga Bd}. We consider $L_\text{fib} = \SI{30}{\kilo\meter}$ of SSMF without fiber attenuation, operated at the C band carrier $\lambda=\SI{1550}{\nano\meter}$. The group velocity dispersion is $\beta_2=\SI{-2.168e-23}{\second^2\per\kilo\meter}$. The sinc pulse and dispersion introduce long memory in the combined channel $g(t)$. The SLD doubles the signal bandwidth to $2B$. The receiver uses a brickwall filter $h(t)$ with bandwidth $2B$. Oversampling $Y(t)$ with $N_\text{sim} = N_\text{os} = 2$ provides sufficient statistics. We approximate $g(t)$ by a discrete filter $g_k$ with $K_g =  151  N_\text{sim} +1$ taps, with a symbol memory of $\widetilde{K}_g = 151$. The filtered and sampled noise $N_k$ is real AWGN. The filter $h(t)$ does not impact the receive signal component, and hence the total  memory~\eqref{eq:total_memory} is $\widetilde{K} = \widetilde{K}_g$.
The average transmit power is $P_\text{tx} = 1/(nT_\mathrm{s}) \,\mathbb{E}[ \lVert X(t) \rVert^2]$ 
and we choose the noise variance $\sigma^2 = 1$, so $\text{SNR}=P_\text{tx}$. The NN parameters are found empirically and are listed in Tab.~\ref{tab:nnparams}.
We compare unipolar $M$-PAM with $\mathcal{A}=\{0,1,\ldots,2^m-1\}$ to bipolar $M$-ASK with $\mathcal{A}=\{\pm1,\pm3,\ldots \pm (2^m-1)\}$; for complex-valued modulation, see~\cite{plabst2024neural}. We use differential phase encoding~\cite{prinz2023successive} before the DAC to help resolve phase ambiguities. 
We plot SIC rates in bits per channel use (bpcu) for the mismatched FBA~\cite{plabst2022achievable}, GS~\cite{prinz2023successive} and NN detectors. Tab.~\ref{tab:complexity} compares the number of multiplications per APP estimate. The FBA must use a mismatched channel memory $\widetilde{N} \ll \widetilde{K}$ because its complexity grows exponentially in $\widetilde{N}$. The mismatched FBA memory is also much smaller than the NN memory. We compute the JDD upper bounds (UBs) of~\cite[Eq.~(45)]{arnoldsimulationmi} via the mismatched FBA with the same memory $\widetilde{N}$~\cite[Sec.~III~C]{plabst2022achievable}.

\begin{figure}
    \centering
    \pgfplotstableread{
    X Y
   -6.0000    0.0754
   -5.0000    0.1169
   -4.0000    0.1792
   -3.0000    0.2705
   -2.0000    0.4004
   -1.0000    0.5785
         0    0.8125
    1.0000    1.1075
    2.0000    1.4643
    3.0000    1.8800
    4.0000    2.3491
    5.0000    2.8640
    6.0000    3.4169
    7.0000    3.9997
    8.0000    4.6053
    9.0000    5.2278
   10.0000    5.8625
   11.0000    6.5056
   12.0000    7.1547
  }{\ASKCovUpperBound}

\pgfdeclarelayer{background}
\pgfdeclarelayer{foreground}
\pgfsetlayers{background,main,foreground}

\input{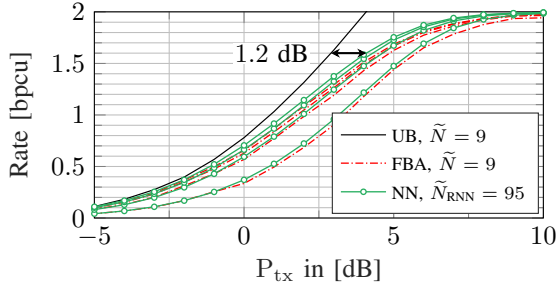}
\begin{tikzpicture}[]

\begin{axis}[%
yminorticks=true,
xmajorgrids,
ymajorgrids,
yminorgrids,
minor x tick num=1,
minor y tick num=4,
grid=both,
legend style={legend cell align=left,  draw=white!15!black, font=\scriptsize,  
},
legend style={at={(0.99,0.02)},anchor=south east,row sep=-1.5pt},
xlabel style={font=\color{white!15!black}},
ylabel style={font=\color{white!15!black}},
axis background/.style={fill=white},
scale only axis,
width=\gridwidth,
height=\gridheight,
xmin=-5,
xmax=10,
xlabel={$\mathrm{P}_{\mathrm{tx}}$ in [dB]},
ymin=0,
ymax=2.001,
ylabel={Rate [bpcu]},
xlabel style={yshift=+2pt},
]

\pgfplotstableread{
X Y
   -5.0000    0.1111
   -4.0000    0.1817
   -3.0000    0.2756
   -2.0000    0.3972
   -1.0000    0.5686
         0    0.7795
    1.0000    1.0320
    2.0000    1.3157
    3.0000    1.6302
    4.0000    1.9724
    5.0000    2.3280
    6.0000    2.6987
    7.0000    3.0672
}{\ASKLoeligerUpperBoundIX}

\addplot [ASK,UB,opacity=0.8,name path global=UB] table {\ASKLoeligerUpperBoundIX};
\addlegendentry{UB, $\widetilde{N} = 9$};

\addplot[ASK,FBA] table [x=power, y=rate, col sep=space]{plots/raw/Q4/TOBI/4-ASK_M=1_alpha=0.00_L=30km_mem=9_n=30000.txt};\addlegendentry{FBA, $\widetilde{N} = 9$}; %

\addplot[ASK,FBA,forget plot] table [x=power, y=rate, col sep=space]{plots/raw/Q4/TOBI/4-ASK_M=2_alpha=0.00_L=30km_mem=9_n=30000.txt};%

\addplot[ASK,FBA,forget plot] table [x=power, y=rate, col sep=space]{plots/raw/Q4/TOBI/4-ASK_M=3_alpha=0.00_L=30km_mem=9_n=30000.txt}; %

\addplot[ASK,FBA,forget plot] table [x=power, y=rate, col sep=space]{plots/raw/Q4/TOBI/4-ASK_M=4_alpha=0.00_L=30km_mem=9_n=30000.txt};%

\addplot[ASK,RNN] table [x=Ptxdb, y=SIC1, col sep=comma]{plots/V1.5/Q4_L30/v1.5,4-ASK,64_128_64_4,lr=5E-04,Tr=64,B=128,S=2,Ls=32,I=2.0E+04,V=1.0E+03,n=6.0E+04,P=151,a=0,L=3.0E+04,Rs=3.5E+10,C=3.1E+04.txt};
\addlegendentry{NN, $\widetilde{N}_\text{RNN}=95$}

\addplot[ASK,RNN,forget plot] table [x=Ptxdb, y=IqXY, col sep=comma]{plots/V1.5/Q4_L30/v1.5,4-ASK,64_128_64_4,lr=5E-04,Tr=64,B=128,S=2,Ls=32,I=2.0E+04,V=1.0E+03,n=6.0E+04,P=151,a=0,L=3.0E+04,Rs=3.5E+10,C=3.1E+04.txt};

\addplot[ASK,RNN,forget plot] table [x=Ptxdb, y=IqXY, col sep=comma]{plots/V1.5/Q4_L30/v1.5,4-ASK,64_128_64_4,lr=5E-04,Tr=64,B=128,S=3,Ls=32,I=2.0E+04,V=1.0E+03,n=6.0E+04,P=151,a=0,L=3.0E+04,Rs=3.5E+10,C=3.1E+04.txt};

\addplot[ASK,RNN,forget plot,name path global=s4ask] table [x=Ptxdb, y=IqXY, col sep=comma]{plots/V1.5/Q4_L30/v1.5,4-ASK,64_128_64_4,lr=5E-04,Tr=66,B=128,S=4,Ls=32,I=2.0E+04,V=1.0E+03,n=6.0E+04,P=151,a=0,L=3.0E+04,Rs=3.5E+10,C=3.1E+04.txt};

\begin{pgfonlayer}{foreground}
\path[name path global=line] (axis cs:\pgfkeysvalueof{/pgfplots/xmin},1.6) -- (axis cs: \pgfkeysvalueof{/pgfplots/xmax},1.6);
\path[name intersections={of=line and UB, name=p1}, name intersections={of=line and s4ask, name=p2}];
\draw[arr] let \p1=(p1-1), \p2=(p2-1) in (p1-1) -- ([xshift=-0.0cm]p2-1) node [left,midway,opacitylabel,font=\normalsize,yshift=0.0cm,xshift=-0.5cm] {%
	\pgfplotsconvertunittocoordinate{x}{\x1}%
	\pgfplotscoordmath{x}{datascaletrafo inverse to fixed}{\pgfmathresult}%
	\edef\valueA{\pgfmathresult}%
	\pgfplotsconvertunittocoordinate{x}{\x2}%
	\pgfplotscoordmath{x}{datascaletrafo inverse to fixed}{\pgfmathresult}%
	\pgfmathparse{\pgfmathresult - \valueA}%
	\pgfmathprintnumber{\pgfmathresult} dB
};
\end{pgfonlayer}
\end{axis}

\end{tikzpicture}%
    \vspace{-12pt}
    \caption{4-ASK, $L_\text{fib}=\SI{30}{\kilo\meter}$. The rates increase with stages $S=1,\ldots,4$; the lower curve of a certain style marks SDD; the upper curve marks $S=4$.} 
    \label{fig:Q4_L=30km}
\end{figure}

Fig.~\ref{fig:Q4_L=30km} and Fig.~\ref{fig:Q8_L=30km} show the SIC rates via the FBA and NN for 4-ASK and 8-ASK, respectively. The rates increase with $S=1,\ldots,4$. We run the FBA for 4-ASK and 8-ASK using $\widetilde{N} = 9$ and $\widetilde{N} = 7$ symbols, respectively. NN-SIC is at least as good as the mismatched FBA-SIC. For 8-ASK, the NN shows large gains over the FBA at medium to high SNRs because of the FBA channel mismatch. The NN substantially reduces the gap to the UB. %
Fig.~\ref{fig:rnn_vs_gibbs_q32_L30} plots the rates of the NNs and bit-wise GS for 32-PAM/ASK and $S=6$ stages. The FBA is infeasible for large constellations. GS runs with $N_\text{par} = 64$ samplers, $\widetilde{N} = 21$ memory, and $N_\text{iter} = 125$ iterations. The performance of the NN and GS is similar at low SNRs. At high SNRs, GS stalls; see~\cite{senst2011rao} and~\cite[Fig.~8]{prinz2023successive}. The NN achieves the maximum rates for 32-PAM/ASK. %
Fig.~\ref{fig:summary} plot the rates for $S=6$ and modulations up to $M=64$. The gains over state-of-the-art unipolar PAM are $\approx \SI{2.1}{dB}$ at 80$\%$ of the maximum rate. Fig.~\ref{fig:complexity} visualizes the algorithmic complexity of Tab.~\ref{tab:complexity}. For 32-PAM/ASK, the NN is 100 times less complex than GS.
\begin{figure}
    \centering
    \pgfplotstableread{
    X Y
 -13.0000    0.0032
  -12.0000    0.0051
  -11.0000    0.0081
  -10.0000    0.0128
   -9.0000    0.0203
   -8.0000    0.0319
   -7.0000    0.0502
   -6.0000    0.0784
   -5.0000    0.1215
   -4.0000    0.1864
   -3.0000    0.2817
   -2.0000    0.4172
   -1.0000    0.6029
         0    0.8469
    1.0000    1.1538
    2.0000    1.5239
    3.0000    1.9535
    4.0000    2.4360
    5.0000    2.9634
    6.0000    3.5270
    7.0000    4.1188
    8.0000    4.7317
    9.0000    5.3600
   10.0000    5.9993
   11.0000    6.6464
   12.0000    7.2990
   13.0000    7.9554
   14.0000    8.6144
   15.0000    9.2752
   16.0000    9.9372
   17.0000   10.6001
   18.0000   11.2635
   }{\ASKCovUpperBound}

\pgfdeclarelayer{background}
\pgfdeclarelayer{foreground}
\pgfsetlayers{background,main,foreground}
\input{plots/plot_settings}

\begin{tikzpicture}[]

\begin{axis}[%
yminorticks=true,
xmajorgrids,
ymajorgrids,
yminorgrids,
minor y tick num=4,
minor x tick num=1,
minor y tick num=4,
grid=both,
legend style={legend cell align=left,  draw=white!15!black, font=\scriptsize,  
},
legend style={at={(0.99,0.02)},anchor=south east,row sep=-1.5pt},
xlabel style={font=\color{white!15!black}},
ylabel style={font=\color{white!15!black}},
axis background/.style={fill=white},
scale only axis,
width=\gridwidth,
height=\gridheight,
xmin=-5,
xmax=15,
xlabel={$\mathrm{P}_{\mathrm{tx}}$ in [dB]},
ymin=0,
ymax=3.001,
ylabel={Rate [bpcu]},
xlabel style={yshift=+2pt},
]

\pgfplotstableread{
    X Y
  -5.000000000000000   0.127659464033783
  -4.000000000000000   0.183545551465475
  -3.000000000000000   0.283199792957735
  -2.000000000000000   0.396136952053306
  -1.000000000000000   0.570516691930347
                   0   0.790497707047351
   1.000000000000000   1.064669330409921
   2.000000000000000   1.355159123505398
   2.999999999999999   1.697875540692198
   4.000000000000000   2.089244923545045
   5.000000000000000   2.519118956528239
   6.000000000000000   2.996732195270778
   7.000000000000000   3.503883175120827
}{\ASKLoeligerUpperBoundXII}
 
\addplot [ASK,UB,black,opacity=0.8,name path global=UB] table {\ASKLoeligerUpperBoundXII};
\addlegendentry{UB, $\widetilde{N} = 7$};

\addplot[ASK,FBA,] table [x=power, y=rate, col sep=space]{plots/raw/Q8/TOBI/8-ASK_M=1_alpha=0.00_L=30km_mem=7.txt}; %
\addlegendentry{FBA, $\widetilde{N} = 7$}; %

\addplot[ASK,FBA,forget plot] table [x=power, y=rate, col sep=space]{plots/raw/Q8/TOBI/8-ASK_M=2_alpha=0.00_L=30km_mem=7.txt};%

\addplot[ASK,FBA,forget plot] table [x=power, y=rate, col sep=space]{plots/raw/Q8/TOBI/8-ASK_M=3_alpha=0.00_L=30km_mem=7.txt};%

\addplot[ASK,FBA,forget plot] table [x=power, y=rate, col sep=space]{plots/raw/Q8/TOBI/8-ASK_M=4_alpha=0.00_L=30km_mem=7.txt};%

\addplot[ASK,RNN] table [x=Ptxdb, y=SIC1, col sep=comma]{plots/V1.5/Q8_L30/v1.5,8-ASK,64_128_128_8,lr=3E-04,Tr=84,B=128,S=2,Ls=32,I=5.0E+04,V=1.0E+03,n=6.0E+04,P=151,a=0,L=3.0E+04,Rs=3.5E+10,C=4.6E+04.txt};
\addlegendentry{NN, $\widetilde{N}_\text{RNN}=115$}

\addplot[ASK,RNN,forget plot] table [x=Ptxdb, y=IqXY, col sep=comma]{plots/V1.5/Q8_L30/v1.5,8-ASK,64_128_128_8,lr=3E-04,Tr=84,B=128,S=2,Ls=32,I=5.0E+04,V=1.0E+03,n=6.0E+04,P=151,a=0,L=3.0E+04,Rs=3.5E+10,C=4.6E+04.txt};

\addplot[ASK,RNN,forget plot] table [x=Ptxdb, y=IqXY, col sep=comma]{plots/V1.5/Q8_L30/v1.5,8-ASK,64_128_128_8,lr=3E-04,Tr=84,B=128,S=3,Ls=32,I=5.0E+04,V=1.0E+03,n=6.0E+04,P=151,a=0,L=3.0E+04,Rs=3.5E+10,C=4.6E+04.txt};

\addplot[ASK,RNN,forget plot,name path global=s4ask] table [x=Ptxdb, y=IqXY, col sep=comma]{plots/V1.5/Q8_L30/v1.5,8-ASK,64_128_128_8,lr=3E-04,Tr=84,B=128,S=4,Ls=32,I=5.0E+04,V=1.0E+03,n=6.0E+04,P=151,a=0,L=3.0E+04,Rs=3.5E+10,C=4.6E+04.txt};

\begin{pgfonlayer}{foreground}
\path[name path global=line] (axis cs:\pgfkeysvalueof{/pgfplots/xmin},2.4) -- (axis cs: \pgfkeysvalueof{/pgfplots/xmax},2.4);
\path[name intersections={of=line and UB, name=p1}, name intersections={of=line and s4ask, name=p2}];
\draw[arr] let \p1=(p1-1), \p2=(p2-1) in (p1-1) -- ([xshift=-0.0cm]p2-1) node [left,midway,opacitylabel,font=\normalsize,yshift=0.0cm,xshift=-0.5cm] {%
	\pgfplotsconvertunittocoordinate{x}{\x1}%
	\pgfplotscoordmath{x}{datascaletrafo inverse to fixed}{\pgfmathresult}%
	\edef\valueA{\pgfmathresult}%
	\pgfplotsconvertunittocoordinate{x}{\x2}%
	\pgfplotscoordmath{x}{datascaletrafo inverse to fixed}{\pgfmathresult}%
	\pgfmathparse{\pgfmathresult - \valueA}%
	\pgfmathprintnumber{\pgfmathresult} dB
};
\end{pgfonlayer}

\end{axis}

\end{tikzpicture}%
    \vspace{-12pt}
    \caption{8-ASK, $L_\text{fib}=\SI{30}{\kilo\meter}$. The rates increase with stages $S=1,\ldots 4$; the lower curve of a certain style marks SDD; the upper curve marks $S=4$.} 
    \label{fig:Q8_L=30km}
\end{figure}
\begin{figure}
    \centering
    \input{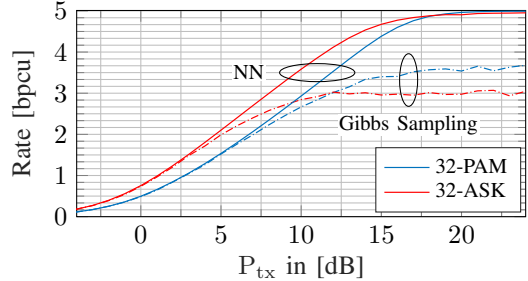}
\begin{tikzpicture}[]

\begin{axis}[%
yminorticks=true,
xmajorgrids,
ymajorgrids,
yminorgrids,
minor x tick num=1,
minor y tick num=5,
grid=both,
legend style={legend cell align=left,  draw=white!15!black, font=\footnotesize,  
at={(0.99,0.02)},anchor=south east,row sep=-1.5pt},
xlabel style={font=\color{white!15!black}},
ylabel style={font=\color{white!15!black}},
axis background/.style={fill=white},
scale only axis,
width=\gridwidth,
height=\gridheight,
xmin=-4,
xmax=24,
xlabel={$\mathrm{P}_{\mathrm{tx}}$ in [dB]},
ymin=0,
ymax=5.01,
ytick={0,1,2,3,4,5},
ylabel={Rate [bpcu]},
xlabel style={yshift=+2pt},
]

\pgfplotstableread[col sep=space]{plots/V1.3/GS/paper__Gibbs__32-PAM__S=6__alpha=0.00__L=30km__mem=21__it=100__par=64__n=40000__burn=25.txt}{\tmpPAM};
\addplot[GS,PAM,forget plot] table [col sep=space,x=power, y expr={ \thisrow{rate} }] {\tmpPAM};

\pgfplotstableread[col sep=space]{plots/V1.3/GS/paper__Gibbs__32-ASK__S=6__alpha=0.00__L=30km__mem=21__it=100__par=64__n=40000__burn=25.txt}{\tmpASK};

\addplot[GS,ASK,forget plot] table [col sep=space,x=power, y expr={\thisrow{rate}}] {\tmpASK};

\pgfplotstableread[col sep=space]{plots/V1.3/GS/paper__Gibbs__32-ring-qam__S=6__alpha=0.00__L=30km__mem=21__it=100__par=64__n=40000__burn=25__m=3-2__phase-offset=0__jlt_spacing=1.txt}{\tmpSQAM};

\addplot[PAM] table [x=Ptxdb, y=IqXY, col sep=comma]{plots/V1.5/Q32_L30/v1.5,32-PAM,100_200_200_200_168_32,lr=4E-05,Tr=120,B=64,S=6,Ls=64,I=1.0E+05,V=7.0E+03,n=8.0E+04,P=151,a=0,L=3.0E+04,Rs=3.5E+10,C=2.3E+05.txt}; 
\addlegendentry{32-PAM}; 

\addplot[ASK] table [x=Ptxdb, y=IqXY, col sep=comma]{plots/V1.5/Q32_L30/v1.5,32-ASK,100_200_200_200_168_32,lr=4E-05,Tr=120,B=64,S=6,Ls=64,I=1.0E+05,V=7.0E+03,n=8.0E+04,P=151,a=0,L=3.0E+04,Rs=3.5E+10,C=2.3E+05.txt}; 
\addlegendentry{32-ASK};

\draw (axis cs:16.7,3.32) ellipse (0.12cm and 0.35cm) node[below,yshift=-0.45cm,align=center,font=\footnotesize,opacitylabel]{Gibbs Sampling};
\draw (axis cs:11.0,3.5) ellipse (0.5cm and 0.12cm) node[left,xshift=-0.68cm,font=\footnotesize,opacitylabel]{NN};

\end{axis}

\end{tikzpicture}%
    \vspace{-12pt}
    \caption{$L_\text{fib}=\SI{30}{\kilo\meter}$ and $S=6$. RNN: $\widetilde{N}_\text{RNN}=169$. Bit-wise GS~\cite{prinz2023successive}: $\widetilde{N}=21$, $N_\text{iter}=125$ and $N_\text{par}=64$.}
    \label{fig:rnn_vs_gibbs_q32_L30}
\end{figure}
\begin{figure}
    \centering
    \input{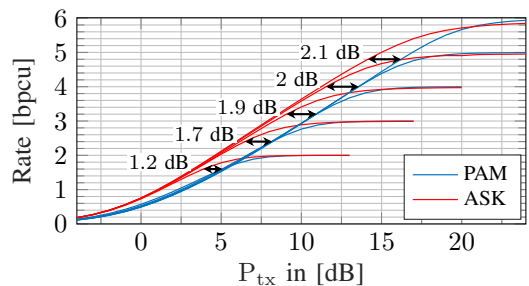}
\begin{tikzpicture}[]

\pgfdeclarelayer{background}
\pgfdeclarelayer{foreground}
\pgfsetlayers{background,main,foreground}

\begin{axis}[%
yminorticks=true,
xmajorgrids,
ymajorgrids,
yminorgrids,
minor x tick num=1,
minor y tick num=4,
grid=both,
legend style={legend cell align=left,  draw=white!15!black, font=\footnotesize,  
at={(0.99,0.02)},anchor=south east,row sep=-1.5pt},
xlabel style={font=\color{white!15!black}},
ylabel style={font=\color{white!15!black}},
axis background/.style={fill=white},
scale only axis,
width=1*\gridwidth,
height=1*\gridheight,
xmin=-4,
xmax=24,
xlabel={$\mathrm{P}_{\mathrm{tx}}$ in [dB]},
ymin=0,
ytick={0,1,...,7},
ymax=6.01,
ylabel={Rate [bpcu]},
xlabel style={yshift=+2pt},
]

\begin{pgfonlayer}{foreground}

\addplot[PAM,name path global=pam4,restrict x to domain=-4:13] table [x=Ptxdb, y=IqXY, col sep=comma]{plots/V1.5/Q4_L30/v1.5,4-PAM,64_128_64_4,lr=5E-04,Tr=120,B=128,S=6,Ls=32,I=2.0E+04,V=1.0E+03,n=6.0E+04,P=151,a=0,L=3.0E+04,Rs=3.5E+10,C=3.1E+04.txt};

\addplot[ASK,restrict x to domain=-4:13,name path global=ask4] table [x=Ptxdb, y=IqXY, col sep=comma,name path global=ask4]{plots/V1.5/Q4_L30/v1.5,4-ASK,64_128_64_4,lr=5E-04,Tr=120,B=128,S=6,Ls=32,I=2.0E+04,V=1.0E+03,n=6.0E+04,P=151,a=0,L=3.0E+04,Rs=3.5E+10,C=3.1E+04.txt};

\addplot[PAM,forget plot,name path global=pam8,restrict x to domain=-4:17] table [x=Ptxdb, y=IqXY, col sep=comma]{plots/V1.5/Q8_L30/v1.5,8-PAM,64_128_128_8,lr=3E-04,Tr=120,B=128,S=6,Ls=32,I=5.0E+04,V=1.0E+03,n=6.0E+04,P=151,a=0,L=3.0E+04,Rs=3.5E+10,C=4.6E+04.txt};

\addplot[ASK,forget plot,name path global=ask8,restrict x to domain=-4:17] table [x=Ptxdb, y=IqXY, col sep=comma]{plots/V1.5/Q8_L30/v1.5,8-ASK,64_128_128_8,lr=3E-04,Tr=120,B=128,S=6,Ls=32,I=5.0E+04,V=1.0E+03,n=6.0E+04,P=151,a=0,L=3.0E+04,Rs=3.5E+10,C=4.6E+04.txt};

\addplot[PAM,forget plot,name path global=pam16,restrict x to domain=-4:20] table [x=Ptxdb, y=IqXY, col sep=comma]{plots/V1.5/Q16_L30/v1.5,16-PAM,84_200_128_128_16,lr=5E-05,Tr=120,B=64,S=6,Ls=64,I=8.0E+04,V=3.0E+03,n=8.0E+04,P=151,a=0,L=3.0E+04,Rs=3.5E+10,C=1.1E+05.txt};

\addplot[ASK,forget plot,name path global=ask16,restrict x to domain=-4:20] table [x=Ptxdb, y=IqXY, col sep=comma]{plots/V1.5/Q16_L30/v1.5,16-ASK,84_200_128_128_16,lr=5E-05,Tr=120,B=64,S=6,Ls=64,I=8.0E+04,V=3.0E+03,n=8.0E+04,P=151,a=0,L=3.0E+04,Rs=3.5E+10,C=1.1E+05.txt};

\addplot[PAM,forget plot,name path global=pam32,restrict x to domain=-4:24] table [x=Ptxdb, y=IqXY, col sep=comma,]{plots/V1.5/Q32_L30/v1.5,32-PAM,100_200_200_200_168_32,lr=4E-05,Tr=120,B=64,S=6,Ls=64,I=1.0E+05,V=7.0E+03,n=8.0E+04,P=151,a=0,L=3.0E+04,Rs=3.5E+10,C=2.3E+05.txt};

\addplot[ASK,forget plot,name path global=ask32,restrict x to domain=-4:24] table [x=Ptxdb, y=IqXY, col sep=comma]{plots/V1.5/Q32_L30/v1.5,32-ASK,100_200_200_200_168_32,lr=4E-05,Tr=120,B=64,S=6,Ls=64,I=1.0E+05,V=7.0E+03,n=8.0E+04,P=151,a=0,L=3.0E+04,Rs=3.5E+10,C=2.3E+05.txt};

\addplot[PAM,forget plot,name path global=pam64,restrict x to domain=-4:24] table [x=Ptxdb, y=IqXY, col sep=comma,]{plots/V1.5/Q64_L30/v1.5,64-PAM,100_300_300_300_240_64,lr=4E-05,Tr=120,B=64,S=6,Ls=100,I=1.0E+05,V=7.0E+03,n=8.0E+04,P=151,a=0,L=3.0E+04,Rs=3.5E+10,C=4.9E+05.txt}; 

\addplot[ASK,forget plot,name path global=ask64,restrict x to domain=-4:24] table [x=Ptxdb, y=IqXY, col sep=comma,]{plots/V1.5/Q64_L30/v1.5,64-ASK,100_300_300_300_240_64,lr=4E-05,Tr=120,B=64,S=6,Ls=100,I=1.0E+05,V=7.0E+03,n=8.0E+04,P=151,a=0,L=3.0E+04,Rs=3.5E+10,C=4.9E+05.txt};

\end{pgfonlayer}

\addlegendentry{PAM}; 
\addlegendentry{ASK};

\begin{pgfonlayer}{foreground}

\path[name path global=line] (axis cs:\pgfkeysvalueof{/pgfplots/xmin},1.6) -- (axis cs: \pgfkeysvalueof{/pgfplots/xmax},1.6);
\path[name intersections={of=line and ask4, name=p1}, name intersections={of=line and pam4, name=p2}];
\draw[arr] let \p1=(p1-1), \p2=(p2-1) in (p1-1) -- ([xshift=-0.0cm]p2-1) node [left,midway,opacitylabel,font=\footnotesize,yshift=0.1cm,xshift=-0.28cm] {%
	\pgfplotsconvertunittocoordinate{x}{\x1}%
	\pgfplotscoordmath{x}{datascaletrafo inverse to fixed}{\pgfmathresult}%
	\edef\valueA{\pgfmathresult}%
	\pgfplotsconvertunittocoordinate{x}{\x2}%
	\pgfplotscoordmath{x}{datascaletrafo inverse to fixed}{\pgfmathresult}%
	\pgfmathparse{\pgfmathresult - \valueA}%
	\pgfmathprintnumber{\pgfmathresult} dB
};

\path[name path global=line] (axis cs:\pgfkeysvalueof{/pgfplots/xmin},2.4) -- (axis cs: \pgfkeysvalueof{/pgfplots/xmax},2.4);
\path[name intersections={of=line and ask8, name=p1}, name intersections={of=line and pam8, name=p2}];
\draw[arr] let \p1=(p1-1), \p2=(p2-1) in (p1-1) -- ([xshift=-0.0cm]p2-1) node [left,midway,opacitylabel,font=\footnotesize,yshift=0.1cm,xshift=-0.28cm] {%
	\pgfplotsconvertunittocoordinate{x}{\x1}%
	\pgfplotscoordmath{x}{datascaletrafo inverse to fixed}{\pgfmathresult}%
	\edef\valueA{\pgfmathresult}%
	\pgfplotsconvertunittocoordinate{x}{\x2}%
	\pgfplotscoordmath{x}{datascaletrafo inverse to fixed}{\pgfmathresult}%
	\pgfmathparse{\pgfmathresult - \valueA}%
	\pgfmathprintnumber{\pgfmathresult} dB
};

\path[name path global=line] (axis cs:\pgfkeysvalueof{/pgfplots/xmin},3.2) -- (axis cs: \pgfkeysvalueof{/pgfplots/xmax},3.2);
\path[name intersections={of=line and ask16, name=p1}, name intersections={of=line and pam16, name=p2}];
\draw[arr] let \p1=(p1-1), \p2=(p2-1) in (p1-1) -- ([xshift=-0.0cm]p2-1) node [left,midway,opacitylabel,font=\footnotesize,yshift=0.1cm,xshift=-0.28cm] {%
	\pgfplotsconvertunittocoordinate{x}{\x1}%
	\pgfplotscoordmath{x}{datascaletrafo inverse to fixed}{\pgfmathresult}%
	\edef\valueA{\pgfmathresult}%
	\pgfplotsconvertunittocoordinate{x}{\x2}%
	\pgfplotscoordmath{x}{datascaletrafo inverse to fixed}{\pgfmathresult}%
	\pgfmathparse{\pgfmathresult - \valueA}%
	\pgfmathprintnumber{\pgfmathresult} dB
};

\path[name path global=line] (axis cs:\pgfkeysvalueof{/pgfplots/xmin},4) -- (axis cs: \pgfkeysvalueof{/pgfplots/xmax},4);
\path[name intersections={of=line and ask32, name=p1}, name intersections={of=line and pam32, name=p2}];
\draw[arr] let \p1=(p1-1), \p2=(p2-1) in (p1-1) -- ([xshift=-0.0cm]p2-1) node [left,midway,opacitylabel,font=\footnotesize,yshift=0.1cm,xshift=-0.28cm] {%
	\pgfplotsconvertunittocoordinate{x}{\x1}%
	\pgfplotscoordmath{x}{datascaletrafo inverse to fixed}{\pgfmathresult}%
	\edef\valueA{\pgfmathresult}%
	\pgfplotsconvertunittocoordinate{x}{\x2}%
	\pgfplotscoordmath{x}{datascaletrafo inverse to fixed}{\pgfmathresult}%
	\pgfmathparse{\pgfmathresult - \valueA}%
	\pgfmathprintnumber{\pgfmathresult} dB
};

\path[name path global=line] (axis cs:\pgfkeysvalueof{/pgfplots/xmin},4.8) -- (axis cs: \pgfkeysvalueof{/pgfplots/xmax},4.8);
\path[name intersections={of=line and ask64, name=p1}, name intersections={of=line and pam64, name=p2}];
\draw[arr] let \p1=(p1-1), \p2=(p2-1) in (p1-1) -- ([xshift=-0.0cm]p2-1) node [left,midway,opacitylabel,font=\footnotesize,yshift=0.1cm,xshift=-0.28cm] {%
	\pgfplotsconvertunittocoordinate{x}{\x1}%
	\pgfplotscoordmath{x}{datascaletrafo inverse to fixed}{\pgfmathresult}%
	\edef\valueA{\pgfmathresult}%
	\pgfplotsconvertunittocoordinate{x}{\x2}%
	\pgfplotscoordmath{x}{datascaletrafo inverse to fixed}{\pgfmathresult}%
	\pgfmathparse{\pgfmathresult - \valueA}%
	\pgfmathprintnumber{\pgfmathresult} dB
};

\end{pgfonlayer}

\end{axis}
\end{tikzpicture}%
    \vspace{-12pt}
    \caption{NN-SIC rates for $L_\text{fib}=\SI{30}{\kilo\meter}$, $S=6$ and $M\in \{4,8,16,32,64\}$.}
    \label{fig:summary}
\end{figure}
\begin{figure}
    \centering
    \input{plots/plot_settings}

\begin{tikzpicture}
	\begin{axis}[
    scale only axis,
    width=1*\gridwidth,
    height=1*\gridheight,
    xmode=log,
    ymode=log,
    xmajorgrids,
    xminorgrids,
    ymajorgrids,
    yminorgrids,
    yminorticks=true,
    grid=both,
    xmin=2^1.7,
    ymin=1E4,
    ymax=1E8,
    ytick={1E4,1E5,1E6,1E7,1E8},
    xtick={4,8,16,32,64},
    xticklabels={4,8,16,32,64},
    xlabel={Constellation size $M$},
    ylabel={Multiplications},
    legend style={legend cell align=left,  draw=white!15!black, font=\footnotesize,  
    at={(0.995,0.99)},anchor=north east,row sep=-1.5pt},
    ]

    \addplot+[FBA,ASK,mark=*,solid,  mark size=1.8pt, mark options={line width=1pt,solid},line width=0.7pt] coordinates { 
    (2^2,1E6) %
    (2^3,16E6) %
    (2^4,4.3E10) %
    };
    \addlegendentry{FBA};

    \addplot+[GS,PAM,mark=square*, solid, mark size=1.6pt, mark options={line width=1pt,solid},line width=0.7pt] coordinates { 
    (2^2,2E5) %
    (2^3,2E5) %
    (2^5,2E7) %
    };
    \addlegendentry{GS};

    \addplot+[RNN,mark=triangle*,solid, mark size=2pt, mark options={line width=1pt,solid},line width=0.7pt] coordinates { 
    (2^2,3E4) %
    (2^3,5E4) %
    (2^4,1E5) %
    (2^5,2E5) %
    (2^6,5E5) %
    };
    \addlegendentry{NN};

	\end{axis}
\end{tikzpicture}
    \vspace{-12pt}
    \caption{Algorithmic complexity~\cite[Tab.~IV]{plabst2024neural}.}
    \label{fig:complexity}
\end{figure}
\section{Conclusions}
\label{sec:conclusions}
We designed time-varying RNN detectors for SIC and showed that they outperform FBA-SIC and GS-SIC with much less complexity. We simulated NN-SIC rates up to \SI{6}{bpcu} with 64-PAM/ASK for a $\SI{30}{\kilo\meter}$ fiber-optic link with a SLD. Bipolar ASK gains up to $\approx\SI{2.1}{dB}$ over state-of-the-art unipolar PAM.
\section*{Acknowledgment}
\noindent The authors wish to thank M. Schädler, S. Calabr\`{o}, D. Lentner Ibanez and the reviewers for useful suggestions.
\clearpage
\bibliographystyle{IEEEtran}
\bibliography{IEEEabrv,nn_dd}

\end{document}